\documentclass[english,aps,prb,twocolumn,amsmath,amssymb,showpacs,superscriptaddress,notitlepage,longbibliography]{revtex4-2}

\usepackage[T1]{fontenc}
\usepackage{color}
\definecolor{page_backgroundcolor}{rgb}{1, 1, 1}
\pagecolor{page_backgroundcolor}
\usepackage{amsmath}
\usepackage{graphicx}
\usepackage{microtype}
\usepackage{bm}
\usepackage{makecell}
\usepackage[normalem]{ulem}
\usepackage{enumitem}
\usepackage[marginal]{footmisc}

\makeatletter
\usepackage{amssymb}
\usepackage{graphicx}
\usepackage[colorlinks=true,linkcolor=blue,anchorcolor=red,citecolor=blue,urlcolor=blue]{hyperref}
\usepackage[caption=false]{subfig}
\usepackage{lipsum}

\makeatother

\usepackage{babel}
\renewcommand{\appendixname}{\MakeUppercase{\appendixname}}
\begin{document}
\global\long\def\figurename{Fig.}

\title{Enhancement of Josephson Supercurrent in a $\pi$-Junction state by Chiral Antiferromagnetism}
\author{Jin-Xing Hou}
\address{Hefei National Laboratory, Hefei, 230088, China}

\author{Hai-Peng Sun}
\affiliation{Shenzhen Key Laboratory of Ultraintense Laser and Advanced Material Technology, Center for Intense Laser Application Technology, and College of Engineering Physics, Shenzhen Technology University, Shenzhen 518118, China}
\affiliation{Institute for Theoretical Physics and Astrophysics, University of W\"urzburg, 97074 W\"urzburg, Germany}
\affiliation{W\"urzburg-Dresden Cluster of Excellence ct.qmat, 97074 W\"urzburg, Germany}

\author{Bj\"orn Trauzettel}
\affiliation{Institute for Theoretical Physics and Astrophysics, University of W\"urzburg, 97074 W\"urzburg, Germany}
\affiliation{W\"urzburg-Dresden Cluster of Excellence ct.qmat, 97074 W\"urzburg, Germany}

\author{Song-Bo Zhang}
\email{songbozhang@ustc.edu.cn}
\address{Hefei National Laboratory, Hefei, 230088, China}
\address{School of Emerging Technology, University of Science and Technology of China, Hefei, 230026, China}

\date{\today}

\begin{abstract}
\vspace{0.8em}

Magnetic order typically disrupts superconductivity, reducing the supercurrent. Here, we show that chiral antiferromagnetism, with non-relativistic spin-split bands and distinctive valley-locked spin texture, can instead significantly enhance Josephson supercurrents. 
This enhancement stems from the emergence of dominant equal-spin triplet pairing and strong fluctuations of singlet pairing in momentum space, both induced by chiral antiferromagnetism. 
We demonstrate these results in Josephson junctions composed of chiral antiferromagnetic metals and conventional superconductors on kagome lattices.  
Furthermore, we show that the enhanced Josephson supercurrent is stabilized in a $\pi$-junction state.
These phenomena persist across a broad energy range and remain stable for different temperatures and junction lengths. 
Our results unveil a previously unexplored mechanism for enhancing supercurrent by strong magnetic order and provide crucial insights into the large Josephson currents observed in Mn$_3$Ge.

\end{abstract}
\maketitle

\section{{Introduction}} Magnetic order is typically considered detrimental to supercurrents because it disrupts superconductivity~\cite{Tinkham2004Book,Golubov04RMP}. The only known exception is in magnetic junctions with antiparallel ferromagnetic bilayers, where phase compensation allows exchange fields to enhance the supercurrent in short junctions at low temperatures~\cite{Bergeret01PRL,Krivoruchko01PRB,Blanter04PRB}. Despite extensive studies~\cite{Barash02PRB,Chtchelkatchev02JETP,Golubov2002JETPL,Bergeret05RMP,Buzdin05RMP,Robinson10PRL,Robinson12SR,eschrig2015spin,linder2015Nphys,Tiira2017NC,HPSun2024arXiv}, however, magnetic enhancement of supercurrents remains rare and poorly understood. Two key questions arise: Can other magnetic orders enhance supercurrents? Which new mechanisms may drive this enhancement? {Addressing these questions is essential for advancing both fundamental understanding and practical applications in superconducting spintronics. 

Recently, unconventional antiferromagnets with non-relativistic spin-split bands, such as altermagnets and chiral antiferromagnets (cAFMs), have attracted increasing interest~\cite{CJWu07PRB,chen2014anomalous,kubler2014non,Rimmler2024NRM,Naka19NC,Ahn19PRB,yuanLD20PRB,Libor20SciAdv,shao2021spin,ma2021multifunctional,Libor22PRX2,Libor22PRXLandscape,Ling24AFM,xiao24PRX,Jiang24PX,Ren24PRX,LHHu25SCPMA}. Several materials, including kagome compounds Mn$_3X$ ($X=$ Sn, Pt, Ir, Ge, Ga)~\cite{liu2018electrical,kiyohara2016giant,nayak2016large,zhang2016giant,liu2017transition,Song2024AFM}, have been identified as cAFMs~\cite{manchon2019current,Jiang24PX,Ren24PRX}. These materials combine key advantages of AFMs~\cite{marti2014room,radu2011transient,qiu2021ultrafast} 
with advanced functionalities~\cite{chen2014anomalous,kubler2014non,martin2008itinerant,nakatsuji2015large,ikhlas2017large,zhang2017strong,higo2018large,kimata2019magnetic}, 
making them highly attractive for spintronic applications~\cite{jungwirth2016antiferromagnetic,baltz2018antiferromagnetic}. Josephson devices based on Mn$_3$Ge thin films have recently been realized, showing unexpectedly large supercurrents~\cite{jeon2021long,jeon2023chiral}. 
These breakthroughs open new avenues for exploring the interplay between magnetism and superconductivity, going beyond traditional collinear AFMs~\cite{Andersen06PRL,Andersen08PRB,Bulaevskii17PRB,Fyhn23PRL,ZhangSB24NC}. 

In this work, we propose a mechanism by which cAFM can substantially enhance the Josephson supercurrent, despite strong spin splitting. This mechanism is driven by dominant triplet pairing and strong singlet pairing fluctuations in momentum space, both of which are uniquely induced by cAFM with valley-locked spin textures and are compatible with band spin splitting. 
Exploiting Josephson junctions made from cAFMs and conventional superconductors on kagome lattices (Fig.~\ref{fig1:mainresult}) as examples, we demonstrate that this mechanism operates generically over a broad energy regime with one Fermi surface per valley and remains robust across various junction lengths and temperatures. Furthermore, it drives the system into a robust $\pi$-junction state. These findings establish a fundamental link between the particular spin-split electronic structure of cAFMs and their capacity to enhance superconducting transport, providing crucial insights into the large supercurrents observed in recent experiments.

{The paper is organized as follows: In Sec.~\ref{sec:general_theory}, we 
formulate the general theory for cAFM-enhanced Josephson supercurrents. Section~\ref{sec:Hamiltonian} introduces the model Hamiltonian for the planar junction with a cAFM sandwiched between two $s$-wave superconductors. We analyze the resulting superconducting pairing correlations in Sec.~\ref{sec:paring_correlations}. The enhanced supercurrent and the emergence of robust $\pi$-junction are studied in Secs.~\ref{sec:enhanced_supercurrent} and \ref{sec:robust_pi_junction}, respectively. Self-consistent calculations are presented in Sec.~\ref{sec:self-consistent-results}. 
Finally, we summarize the main conclusions in Sec.~\ref{sec:discussion}.}

\

\

\begin{figure*}[t]
\centering
\includegraphics[width=0.98\linewidth]{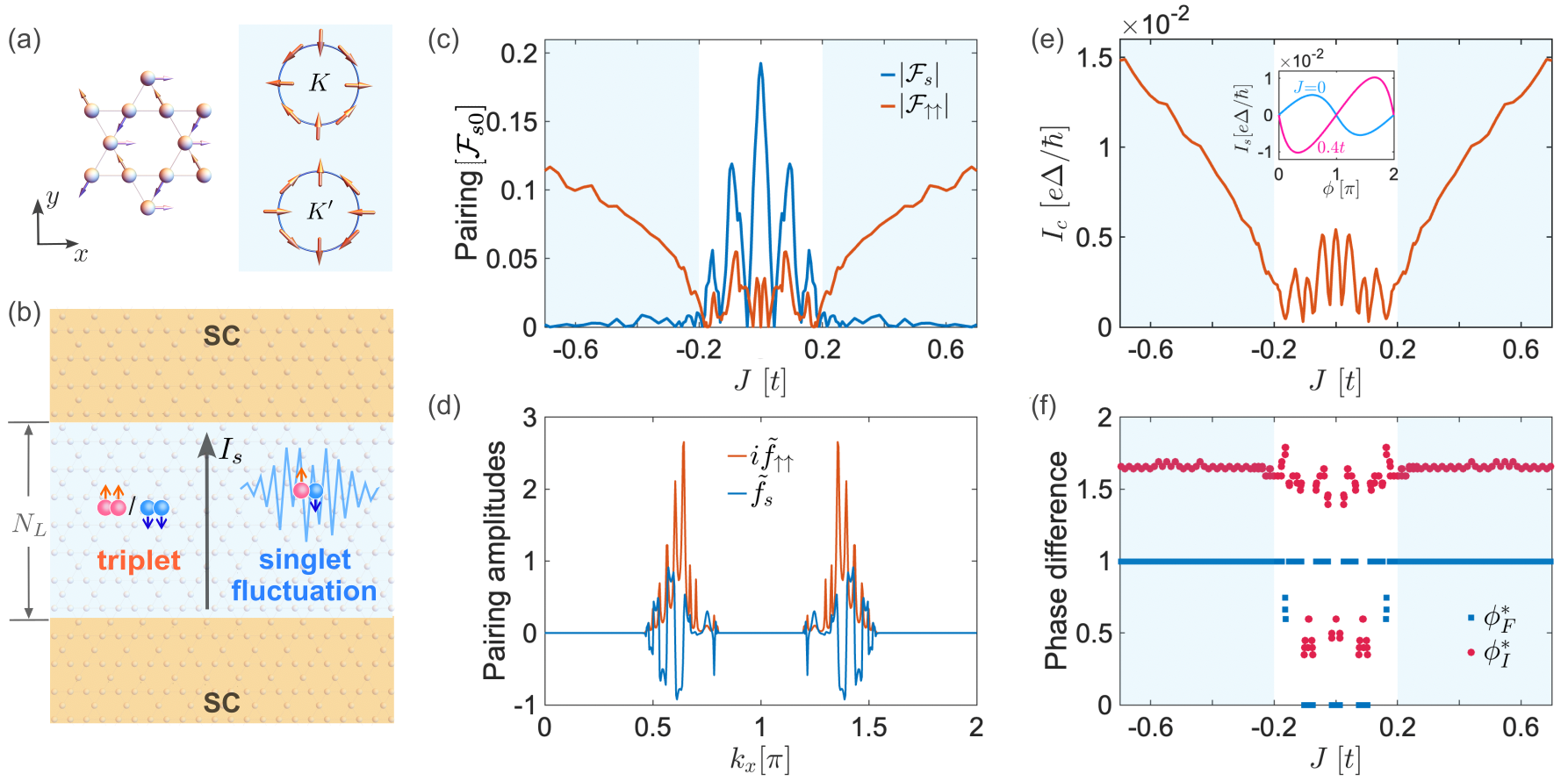}
\caption{(a) Left: cAFM on kagome lattices, where the arrows indicate the local magnetic moments. Right: Fermi surfaces at valleys $K$ and $K'$ with valley-dependent spin textures for $|\mu_{\text{AFM}}|<|J|$. 
(b) Schematic of the Josephson junction, where the yellow regions represent the superconducting leads, while the cyan region is the cAFM with $N_L$ layers (in units of $\sqrt{3}a$). 
(c) Net spin-singlet $|\mathcal{F}_s|$ (blue) and spin-triplet $|\mathcal{F}_{\uparrow\uparrow(\downarrow\downarrow)}|$ (purple) pairing amplitudes (in units of $\mathcal{F}_{s0}$) at the junction center ($y=N_L/2$) as functions of $J$. Here, $\mathcal{F}_{s0}$ is the singlet pairing amplitude in the bulk superconductor. For illustration, $\omega_0=0.1 k_B T_c$ is used. (d) $k_x$-resolved singlet and triplet pairing amplitudes, $i\tilde{f}_{\uparrow\uparrow}$ and $\tilde{f}_s$, at the junction center. (e) Maximum supercurrent $I_c$ (in units of $e\Delta/\hbar$) as a function of $J$ for temperature $T=0.02T_c$. Insets show the current-phase relations at $J=0$ (blue) and $0.4t$ (purple), respectively.  (f) Phase position $\phi^*_{I}$ (purple, in units of $\pi$) for $I_c$ and $\phi^*_{F}$ (blue, in units of $\pi$) for the lowest free energy as functions of $J$. (c)-(f) are calculated with the self-consistently determined superconducting order parameter. 
Other parameters: $\mu_S=2t$, $\mu_{\text{AFM}}=0.2t$, $N_L=60$, and $U=0.65t$ which yields $\Delta=0.023t$ and $k_BT_c = 0.013t$. }
\label{fig1:mainresult}
\end{figure*}

\section{{General theory of cAFM-enhanced Josephson effect}}\label{sec:general_theory}
In ferromagnetic Josephson junctions, the exchange field causes spin splitting in the electronic bands and disrupts the coherent propagation of singlet Cooper pairs. Phase compensation can only occur when antiparallel ferromagnetic domains of equal length are present. Additionally, magnetic stray fields often emerge, further undermining superconductivity. Consequently, supercurrent is typically suppressed compared to the non-magnetic case. 

The situation is fundamentally distinct in cAFMs with spin-split bands and noncollinear magnetic order [see a sketch in Figs.~\ref{fig1:mainresult}(a) and \ref{fig1:mainresult}(b)]. In the strong spin-split regime, each valley (at the $K$ and $K'$ points) hosts one Fermi surface with opposite in-plane spin textures. Consequently, cAFMs exhibit half-metal-like behavior despite having zero net magnetization, where an electron with a given spin and momentum cannot pair with an electron of opposite spin and momentum, leading to suppression of net singlet pairing. However, due to the intrinsic noncollinear magnetic nature, pronounced equal-spin triplet Cooper pairs are efficiently converted from singlet pairs at the interface, which dominate superconducting transport [Fig.~\ref{fig1:mainresult}(c)]. 
Meanwhile, singlet pairing can be finite for individual momenta but exhibits strong fluctuations in momentum space, although its net value becomes vanishingly small across the junction [Fig.~\ref{fig1:mainresult}(d)].
These emergent pairing channels--equal-spin triplet pairing and singlet fluctuations--become more pronounced with increasing cAFM strength. They facilitate robust Cooper-pair propagation across the junction without being disrupted by strong spin splitting. Moreover, unlike ferromagnetic junctions, the cAFM junction has zero net magnetization and avoids stray fields, ensuring better compatibility with superconductivity. Therefore, we expect enhanced supercurrents due to the cAFM order [Figs.~\ref{fig1:mainresult}(e) and \ref{fig1:mainresult}(f)]. 
Note that this mechanism is unique to cAFM with distinctive valley-dependent spin textures and is absent in junctions with collinear antiferromagnets or ferromagnets.}  

\section{{Model Hamiltonian}}\label{sec:Hamiltonian}
To illustrate the above essential physics, we consider planar Josephson junctions composed of a cAFM metal and two conventional $s$-wave superconductors on kagome lattices [Fig.~\ref{fig1:mainresult}(b)]. 
We assume translational symmetry along the interfaces, i.e., $x$ direction. Thus, the momentum $k_x$ remains a good quantum number. The  junction can be described by  $H=H_0+H_{\text{cAFM}}+H_\Delta$, where $H_0$ is the prototype tight-binding kagome model 
\begin{equation} \label{eq:kagome-model}
    H_0=  \sum_j \big\{\Psi_{j}^\dag \mathcal{H}(k_x) \Psi_{j} + \big[\Psi_{j}^\dag \mathcal{T}(k_x) \Psi_{j+1}+\mathrm{h.c.}\big]\big\},
\end{equation}
where $\Psi_j^\dagger = (c_{j,1,\uparrow}^\dagger, c_{j,2,\uparrow}^\dagger, c_{j,3,\uparrow}^\dagger,c_{j,1,\downarrow}^\dagger, c_{j,2,\downarrow}^\dagger, c_{j,3,\downarrow}^\dagger)$,  
$c_{j,\nu,s}^\dag$ creates an electron with spin $s\in\{\uparrow,\downarrow\}$ at sublattice $\nu$ of the $j$-th layer {  of unit cells along $y$-direction}, 
$\mathcal{H}(k_x) = t s_0 \otimes h(k_x)$ and $\mathcal{T}(k_x)=t s_0 \otimes T^+(k_x)$ are the local terms for each layer and hopping matrix between neighboring layers, respectively. $s_0$ is the unit matrix, $\bm{s} = (s_x,s_y,s_z)$ is the Pauli matrix vector for spin, $t$ is the hopping amplitude between neighboring sites, $h(k_x)$ and $T^+(k_x)$ are specified in Appendix \ref{sec:bandstructure}. Without superconductivity and magnetism, the model hosts spin-degenerate Dirac electrons at the $K$ and $K'$ points (valleys) in the Brillouin zone, with energies near the band center. We set the Dirac points at zero energy and the lattice constant to be $a=1$. 

The $H_{\text{cAFM}}$ term  describes the cAFM order in the middle region with $N_L$ layers: 
\begin{equation}\label{eq:H_magnetic_order}
 H_{\text{cAFM}} = \sum_{1\leqslant j\leqslant N_L} \sum_{\nu=1,2,3} \Psi^\dagger_{j,\nu}\ \bm{m}_{\nu} \cdot \bm{s}\ \Psi_{j,\nu}, 
\end{equation}
where $\bm{m}_\nu = J (\cos{\theta_\nu},\sin{\theta_\nu},0)$ is the magnetic moment at sublattice $\nu$ with strength $J$ and direction $\theta_\nu = 2(3-\nu)\pi/3$.  $\bm{m}_\nu$ points at angles of 120$^\circ$ with respect to each other, as observed in the Mn$_3X$ compounds [Fig.~\ref{fig1:mainresult}(a)]. In real materials such as Mn$_3$Ge and Mn$_3$Ga, the strength $J$ is comparable to the hopping energy $t$~\cite{zhang2017strong,Song2024AFM}. This results in strong spin splitting in the band structure of the cAFM~\cite{chen2014anomalous}. Interestingly, for { Fermi energies} $|\mu_{\text{AFM}}|<|J|$ ({ measured from the band center}), one Fermi surface forms at each valley, with an even-parity valley-dependent spin texture~\cite{SBZhang25arXiv-a}.

Within mean-field theory, the pairing term in the superconducting leads (for $j\leqslant 0$ and $j\geqslant N_L+1$) can be written as 
\begin{equation}
\label{eq:H_SC_pairing} 
 H_{\Delta} = \Big(\sum_{j\leqslant 0}  +\sum_{j\geqslant N_L+1} \Big) \Delta_j \Psi_{j,\uparrow}^\dag (\Psi_{j,\downarrow}^\dag)^T + \mathrm{h.c.} ,
\end{equation}
where $\Delta_j$ is a position-dependent pairing potential. It can be solved self-consistently from the attractive Hubbard interaction $ H_{U} = - \sum_{j} U_j c_{j,\nu,\uparrow}^\dag
c_{j,\nu,\uparrow} c_{j,\nu,\downarrow}^\dag c_{j,\nu,\downarrow}$, with a uniform strength $U_j=U$ in the superconducting regions~\cite{Awoga19PRL,Black-Schaffer08PRB,Setiawan2019PRB}. We consider a large { Fermi energy} $\mu_S$ ($\sim 2t$) in the superconductors, yielding a large Fermi surface at the $\Gamma$ point, as in conventional superconductors. This creates a large Fermi-level mismatch ($\mu_S\gg\mu_{\text{AFM}}$) between the cAFM and superconductors, leading to step-like variations of $\Delta_j$ at the interfaces~\cite{Black-Schaffer08PRB,Breunig2021PRB,Setiawan2019PRB}. To elucidate the main physics and for simplicity, we first assume a step-like profile for $\Delta_j$, i.e., $|\Delta_j|=\Delta_0$, with a phase difference $\phi$ across the junction. Later, we demonstrate that our key results are valid employing a self-consistent treatment. { They also apply qualitatively to junctions oriented along other directions.}

\begin{figure}[t]
\centering
\includegraphics[width=1\linewidth]{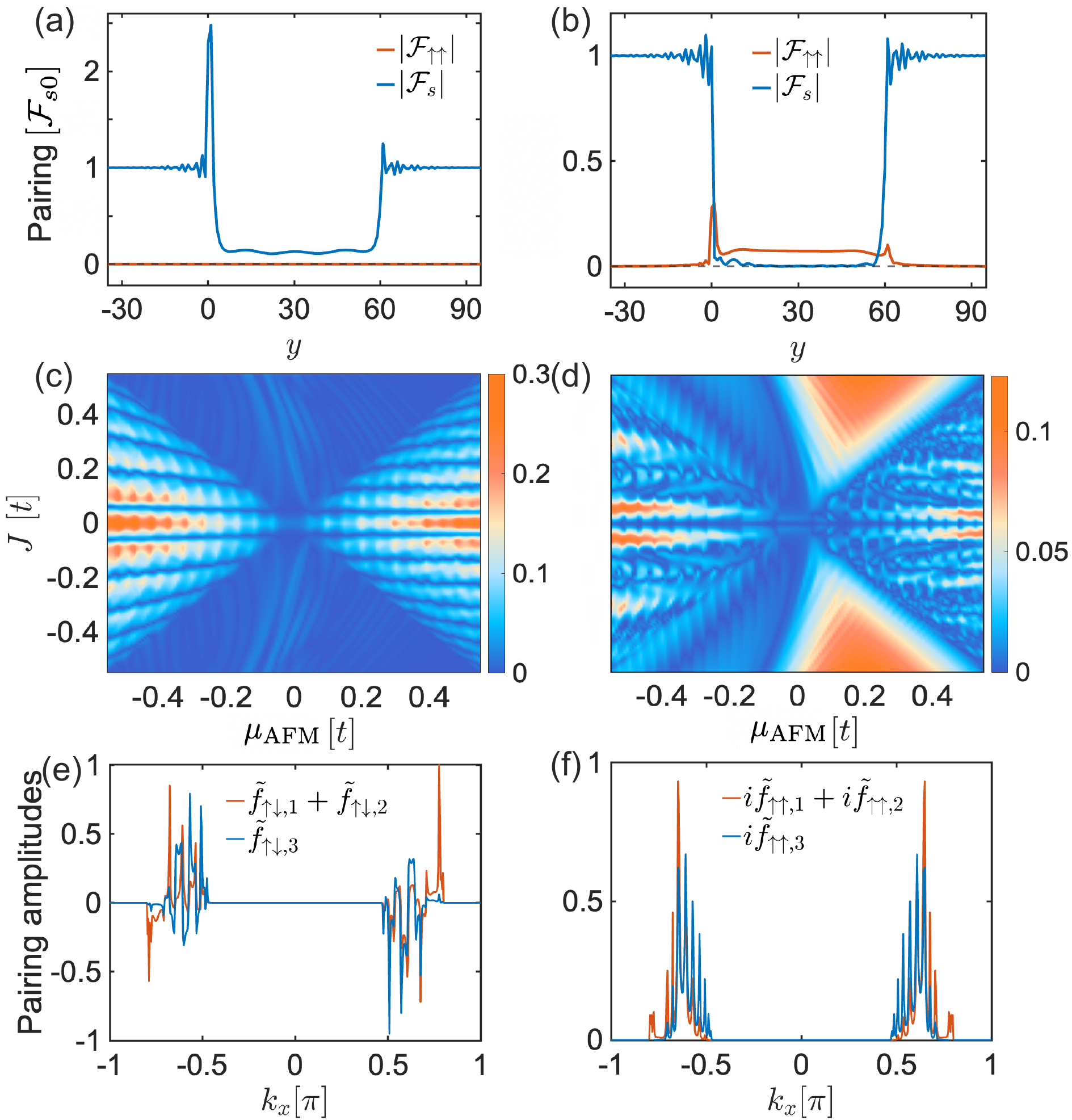} 
\caption{(a) Singlet $|\mathcal{F}_s|$ (blue) and equal-spin triplet  $|\mathcal{F}_{\uparrow\uparrow(\downarrow\downarrow)}|$ (red) pairing amplitudes in the junction without cAFM order ($J=0$) for $\mu_{\text{AFM}}=0.2t$. 
(b) Same as (a) but in presence of cAFM ($J=0.4t$).  
(c) $|\mathcal{F}_s|$ at the junction center ($y=N_L/2$) as a function of $J$ and $\mu_{\text{AFM}}$. 
(d) $|\mathcal{F}_{\uparrow\uparrow(\downarrow\downarrow)}|$ at the junction center as a function of $J$ and $\mu_{\text{AFM}}$. (e, f) $k_x$ and sublattice-resolved singlet and triplet pairing amplitudes at the junction center for $J=0.4t$ and $\mu_{\text{AFM}}=0.2t$. Other parameters: $\Delta_0 = 0.02t$, $\omega_0=0.1\Delta_0$, $N_L=60$, and $\mu_S=2t$.
}
\label{fig:PairingCorrelation}
\end{figure}

\section{{Pairing correlations}}\label{sec:paring_correlations}
The unique valley-locked spin texture of the cAFM profoundly affects the pairing correlations across the junction, resulting in pronounced triplet pairing and strong fluctuations of singlet pairing. To illustrate this, we calculate
local pairing correlations from the anomalous Green function $\hat{F}_{eh}$. In spin space, $\hat{F}_{eh}$ can be decomposed into singlet and triplet components as~\cite{Breunig19PRB}
\begin{equation} \label{eq-Feh}
 \hat{F}_{eh} (y,k_x,i\omega_n) 
 = -is_y \sum_{\alpha={0,x,y,z}} \tilde{\bm f}_\alpha(y,k_x,i\omega_n) s_\alpha, 
\end{equation} 
where $y$ is the layer position, $\omega_n=(2n + 1)\pi k_BT$ are Matsubara frequencies with integers $n$, Boltzmann constant $k_B$ and temperature $T$. The component $\tilde{\bm f}_s\equiv \tilde{\bm f}_0$ corresponds to the singlet pairing amplitudes, while $\tilde{\bm f}_z$ and $\tilde{\bm f}_{\uparrow\uparrow/\downarrow\downarrow}= \mp {\tilde{\bm f}}_x - i {\tilde{\bm f}}_y$ are triplet pairing amplitudes. They are $3\times 3$ matrices in sublattice space.  

Figures~\ref{fig:PairingCorrelation}(a) and \ref{fig:PairingCorrelation}(b) present the net pairing amplitudes, i.e., $\mathcal{F}_\alpha = \int {dk_x}\text{Tr} [{\tilde{\bm f}}_\alpha]/{2\pi}$, in the junction, integrating over momentum space and summing over the sublattices in each unit cell. For illustration, $\omega_0=0.1\Delta_0$ is used, but the results are qualitatively the same for any finite $\omega_n$. In absence of cAFM ($J=0$), only singlet pairing $\mathcal{F}_s$ is present, as expected [Fig.~\ref{fig:PairingCorrelation}(a)]. Upon introducing cAFM, two equal-spin triplet pairing correlations of opposite amplitudes $\mathcal{F}_{\uparrow\uparrow}=-\mathcal{F}_{\downarrow\downarrow}$ emerge. The singlet pairing is even in frequency, while the triplet pairing is odd, as required by the Pauli principle~\cite{BalatskyPhysRevB1992, Tanaka2007PRB,Tanaka2011symmetry}. Moreover, the triplet amplitudes have $\pm \pi$ phase difference relative to the singlet one.  {  The asymmetric profile at the junction interfaces arises from a structural difference between the upper and lower interfaces of the kagome lattice.} 

Remarkably, for $|J|$>$|\mu_{\text{AFM}}|$, while  $\mathcal{F}_s$ is rapidly suppressed in the cAFM, pronounced $\mathcal{F}_{\uparrow\uparrow}$ and $\mathcal{F}_{\downarrow\downarrow}$ dominate across the junction [Fig.~\ref{fig:PairingCorrelation}(b)]. 
This is further confirmed by Figs.~\ref{fig:PairingCorrelation}(c) and \ref{fig:PairingCorrelation}(d), where we calculate the phase diagrams of $|\mathcal{F}_s|$ and $|\mathcal{F}_{\uparrow\uparrow(\downarrow\downarrow)}|$ at the junction center ($y=N_L/2$) as functions of $J$ and $\mu_{\text{AFM}}$. In this regime, the triplet pairing grows monotonically as $J$ ($<|t|$) increases. 
Moreover, while net singlet pairing is fully suppressed in the junction, strong singlet fluctuations emerge in momentum space. Figures~\ref{fig:PairingCorrelation}(e) and \ref{fig:PairingCorrelation}(f) present the $k_x$-resolve pairing amplitudes ${\tilde f}_{\alpha,\nu}$ at the junction center. We find that these amplitudes are most pronounced around the valley points ($k_x = \pm \pi/\sqrt{3}$) where the Fermi surfaces are located. Interestingly, while the triplet pairing retains the same sign in $k_x$, the singlet pairing fluctuates strongly around zero, even though its net sum is zero. These fluctuations are comparable in magnitude to the triplet pairing and become faster for stronger $J$ and longer junctions, as shown in Appendix \ref{Sec:Analysis}. 

\begin{figure}[t]
    \centering
    \includegraphics[width=1\linewidth]{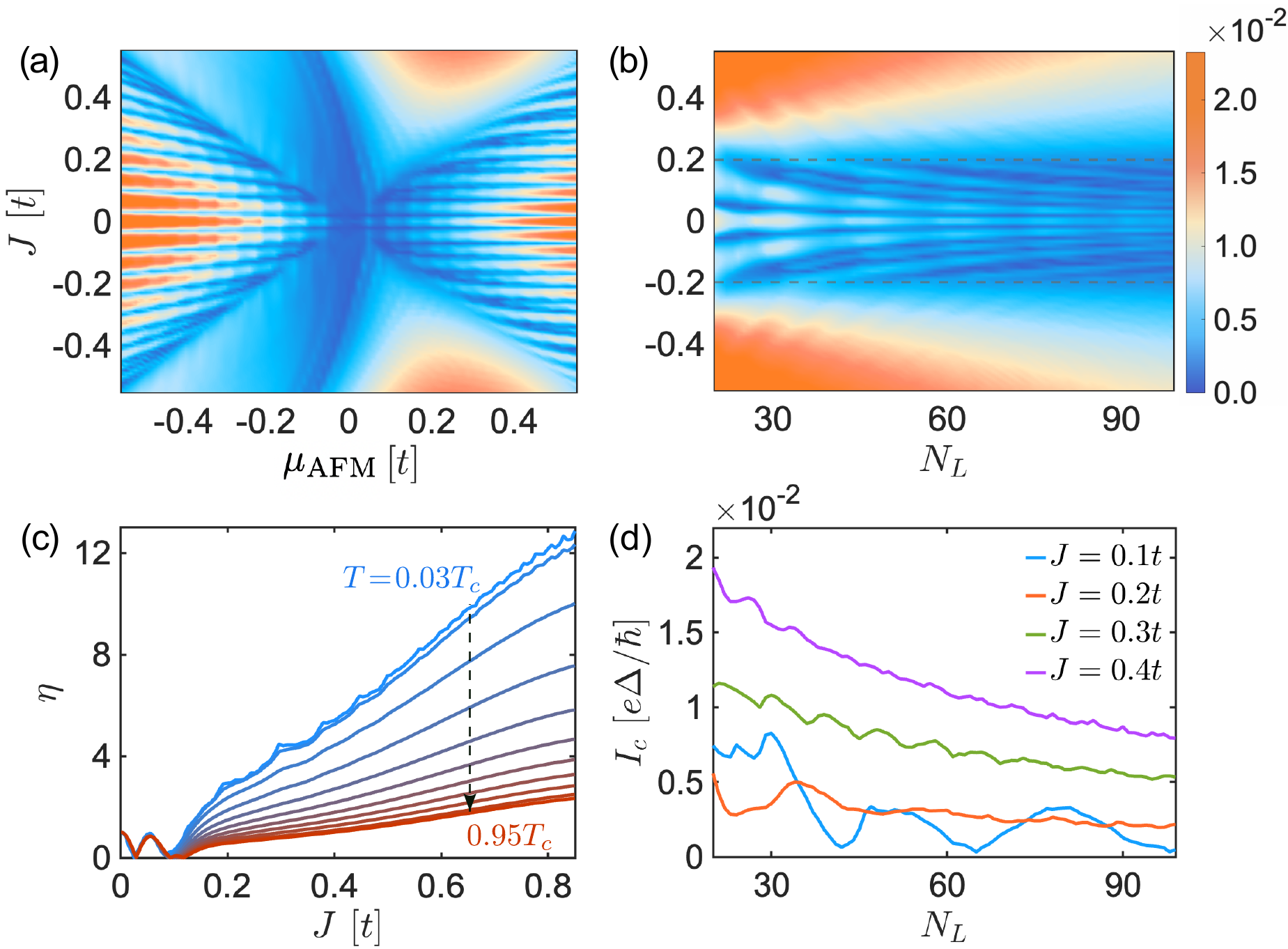}
    \caption{(a) Maximum supercurrent $I_c$ (in units of $e\Delta/\hbar$) as a function of $J$ and $\mu_{\text{AFM}}$ for $N_L=60$. (b) $I_c$ as a function of $J$ and $N_L$ for $\mu_{\text{AFM}}=0.2t$. (c) Enhancement ratio $\eta=I_c/I_c(J=0)$ as a function of $J$ for $N_L=50$, $\mu_{\text{AFM}}=0.1t$ and increasing $T$ (i.e., $T=0.03T_c$, $0.1T_c$, $\cdots$, $0.9T_c$, and $0.95T_c$, where $T_c=0.57\Delta$). (d) $I_c$ as a function of $N_L$ for $\mu_{\text{AFM}}=0.2t$ and various $J$. $k_BT=0.02\Delta$ in (a,b,d), $\Delta=0.02t$ and 
    $\mu_S=2t$ for all panels.}
    \label{fig:enhance}
\end{figure}

\section{{Enhanced supercurrent}}\label{sec:enhanced_supercurrent}
The induced dominant triplet pairing and fluctuations of singlet pairing directly contribute to and substantially enhance Josephson supercurrents. The supercurrent can be obtained as a response of the junction free energy $F$ to the change of $\phi$, $I_s=({2e}/{\hbar}){\partial F(\phi)}/{\partial \phi}$, where $e$ is the elementary charge and $\hbar$ the reduced Planck constant~\cite{Tinkham2004Book}.  
To better understand the connection between pairing correlations and supercurrents, it is instructive to apply the Green function technique~\cite{Sancho1985JPFMP,Asano01prb,SBZhang20PRB}. Both methods yield identical results, as confirmed in Appendix \ref{sec:compare_two _methods}. 
In the Green function method, the supercurrent $I_s$ can be decomposed into two parts: one from triplet pairing and the other from singlet pairing. In the weak-coupling regime, it is approximately $I_s \approx (I_{s,c}+I_{t,c})\sin \phi$, where
\begin{subequations}
\begin{align}
I_{s,c} = & \; \dfrac{2e}{\pi\hbar} k_BT t^2  \sum_{\omega_n} \int {dk_x}   
\big[(\tilde{f}_{\downarrow\uparrow,1}+\tilde{f}_{\downarrow\uparrow,2})\tilde{f}_{\uparrow\downarrow,3}' \big], \label{eq:5a} \\
I_{t,c} = & \; \dfrac{2e}{\pi\hbar} k_BT t^2 \sum_{\omega_n} \int {dk_x}   
\big[(\tilde{f}_{\uparrow\uparrow,1} + \tilde{f}_{\uparrow\uparrow,2})\tilde{f}_{\uparrow\uparrow,3}'
  \big]. \label{eq:5b}
\end{align}
\end{subequations}
Here, $\tilde{f}_{\uparrow\downarrow,\nu}$ and $\tilde{f}_{\uparrow\uparrow,\nu}$ { (corresponding to the diagonal elements of $\tilde{\bm{f}}_{\alpha}$)} are singlet and triplet pairing correlations at sublattice $\nu$ at the junction center for $\phi=0$. They satisfy the symmetry relation
$\tilde{f}_{\sigma\sigma',\nu}'(k_x)=-\tilde{f}_{\sigma\sigma',\nu}^*(-k_x)$. 
From Eq.~\eqref{eq:5a}, we find that even though the net singlet pairing vanishes ($\mathcal{F}_s=0$), its strong fluctuations can still significantly contribute to $I_s$.

Figure~\ref{fig:enhance}(a) presents the maximum supercurrent $I_c=\max[I_s(\phi)]$ as a function of $J$ and $\mu_{\text{AFM}}$ at low temperature ($k_BT=0.02\Delta_0$) for junction length $N_L=60$. Strikingly, this diagram closely mirrors that of pairing amplitudes [Figs.~\ref{fig:PairingCorrelation}(c) and \ref{fig:PairingCorrelation}(d)].
For $|J|>|\mu_{\text{AFM}}|$ where there is one Fermi surface at each valley, $I_c$ increases with $|J|(<|t|)$ and can be substantially enhanced by $J$ for $\mu_{\text{AFM}}>0$. The enhancement ratio, $\eta=I_c(J)/I_c(J=0)$, is larger for smaller $\mu_{\text{AFM}}$ and stronger $J$ [Fig.~\ref{fig:enhance}(c)]. It can exceed a factor $10$ at $J=0.8t$ for $\mu_{\text{AFM}}=0.1t$. 
In Fig.~\ref{fig:enhance}(b), we further calculate $I_c$ as a function of $J$ and $N_L$ for fixed $\mu_{\text{AFM}}(=0.2t)$. As $N_L$ increases, $I_c$ decays overall with oscillations for any $J$. Moreover, for $0<|J|<\mu_{\text{AFM}}$, $I_c$ periodically drops to zero, indicating $0$-$\pi$ transitions [Fig.~\ref{fig:enhance}(d)]. 
In contrast, for $J=0$ and $|J|>\mu_{\text{AFM}}$, $I_c$ decays more smoothly with slight oscillations. Most interestingly, $I_c$ is significantly enhanced by $|J|>\mu_{\text{AFM}}$ for different $N_L$, compared to the nonmagnetic case ($J=0$). 
This magnetic enhancement persists even at elevated temperatures up to the superconducting critical temperature $T_c=0.57\Delta$ [Fig.~\ref{fig:enhance}(c)].

\begin{figure}[t]
    \centering
    \includegraphics[width=1\linewidth]{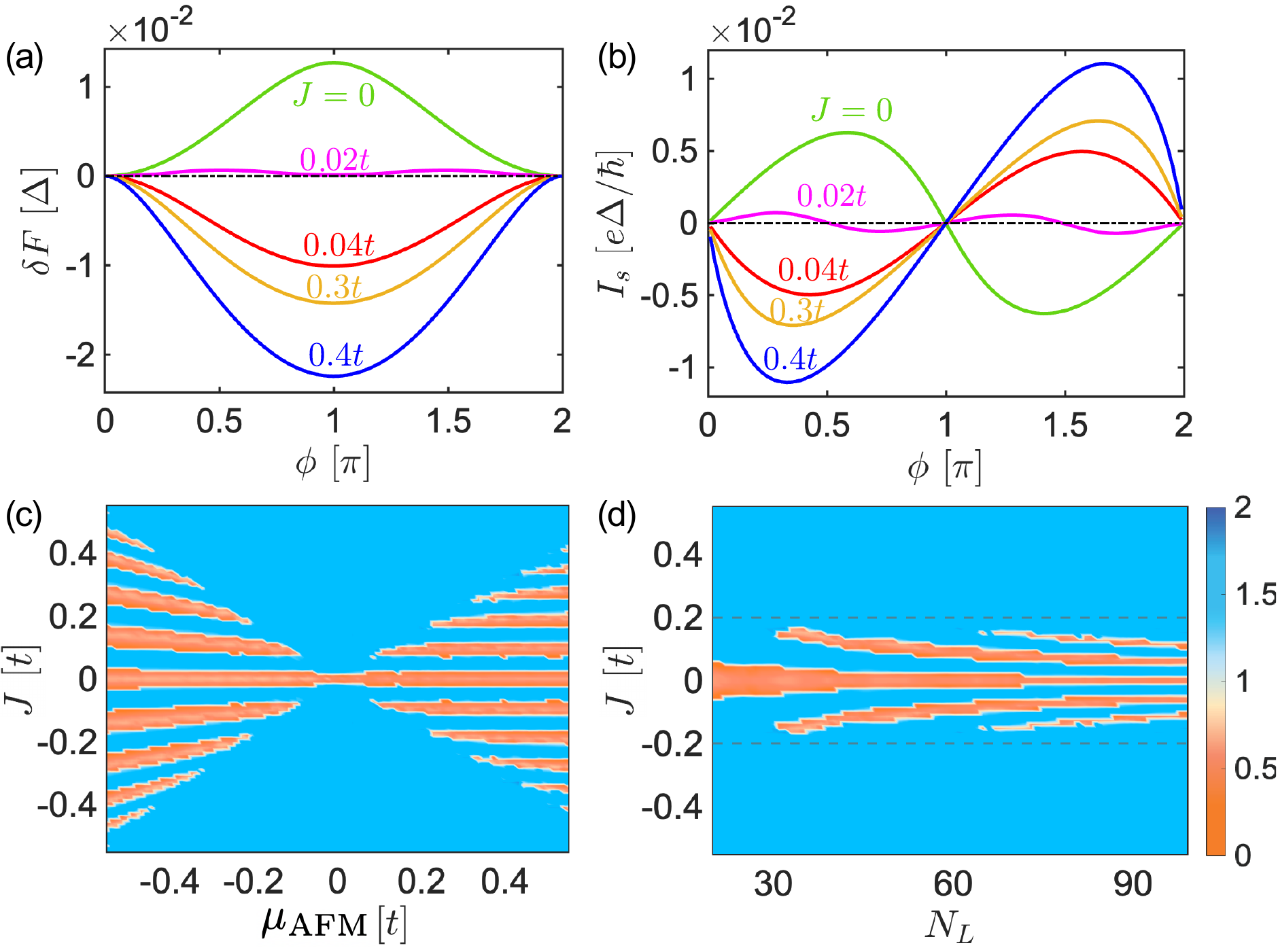}
    \caption{(a) Free energy $\delta F(\phi)= F(\phi)-F(\phi=0)$, measured relative to the value at $\phi=0$, as a function of phase difference $\phi$ for various $J$. 
    (b) CPR $I_s(\phi)$ for various $J$, corresponding to (a). 
    (c) Phase position $\phi^*_I$ (in units of $\pi$) of $I_c$ as a function of $J$ and $\mu_{\text{AFM}}$. 
    (d) $\phi^*_I$ as a function of $J$ and $N_L$ for $\mu_{\text{AFM}}=0.2t$. 
    Other parameters are the same as Fig.~\ref{fig:PairingCorrelation}.}
    \label{fig:pi_junction}
\end{figure}

\section{{{Robust $\pi$-junction}}}\label{sec:robust_pi_junction}
In the regime $|J|>|\mu_{\text{AFM}}|$ with dominant triplet pairing and strong singlet fluctuations, the junction is stabilized in a $\pi$ state, characterized by an intrinsic $\pi$ phase difference in the ground state. This is evident from the calculated free energy $F(\phi)$ for varying $J$ [Fig.~\ref{fig:pi_junction}(a)]. 
For $|J|<|\mu_{\text{AFM}}|$, the phase position of the lowest free energy, $\min[F(\phi)]$, can be either at $\phi_F^*=0$ or $\pi$, depending on the specific values of $\mu_{\text{AFM}}$, $J$ and $N_L$. Thus, the junction can be either a 0-junction or a $\pi$-junction.  
In contrast, for $|J|>|\mu_{\text{AFM}}|$, the lowest free energy always shifts to be at $\phi_F^*=\pi$, confirming the presence of a $\pi$-junction. 
The corresponding current-phase relations are shown in Fig.~\ref{fig:pi_junction}(b). In $0$-junctions, the maximum current $I_c$ occurs in the interval $\phi^*_{I}\in[0,\pi]$, whereas in $\pi$-junctions, $I_c$ is found in $\phi^*_{I}\in[\pi,2\pi]$.

In Fig.~\ref{fig:pi_junction}(c), we calculate $\phi^*_{I}$ as a function of $J$ and $\mu_{\text{AFM}}$ with $N_L=60$. For $|J|>|\mu_{\text{AFM}}|$, $\phi^*_{I}$ is consistently pinned in the phase interval $\phi^*_{I}\in[\pi,2\pi]$, while for $|J|<|\mu_{\text{AFM}}|$, $\phi^*_{I}$ oscillates between values greater than and less than $\pi$, showing 0-$\pi$ transitions.
Figure~\ref{fig:pi_junction}(d) presents $\phi^*_{I}$ as a function of $J$ and $N_L$ for $\mu_{\text{AFM}}=0.2t$. It demonstrates that $\phi^*_{I}$ remains larger than $\pi$ for $|J|>\mu_{\text{AFM}}$ regardless of $N_L$, while, for $|J|<\mu_{\text{AFM}}$, $\phi^*_{I}$ shifts between greater than and less than $\pi$ depending on $J$ and $N_L$. These results further support that a $\pi$-junction is always induced and enhanced by $J$ in the regime $|J|>|\mu_{\text{AFM}}|$, irrespective of the specific model parameters. 
Finally, we note that our $\pi$-junction emerges without net magnetization, in contrast to junctions with antiparallel ferromagnets where no $\pi$ state can be observed~\cite{Krivoruchko01PRB,Blanter04PRB}.

\section{{Self-consistent results}}\label{sec:self-consistent-results}
Above, we have demonstrated the key results under the assumption of constant pairing potentials in the superconducting leads in the presence of large Fermi-level mismatch. However, we emphasize that the main results are not restricted to such assumptions. In Figs.~\ref{fig1:mainresult}(c)-(f), we perform self-consistent mean-field calculations of the pairing potential and supercurrent using large computationally feasible system sizes, following the well-established approach developed in Refs.~\cite{Black-Schaffer08PRB,Awoga19PRL,Setiawan2019PRB}. In our calculation, we adopt $U=0.65t$, which yields a bulk pairing potential $\Delta \approx 0.023t$, closely matching the previously assumed constant pairing potential $\Delta_0=0.02t$ (see Appendix \ref{sec:AP_self_consistent}).

\begin{figure}[t]
    \centering
    \includegraphics[width=1\linewidth]{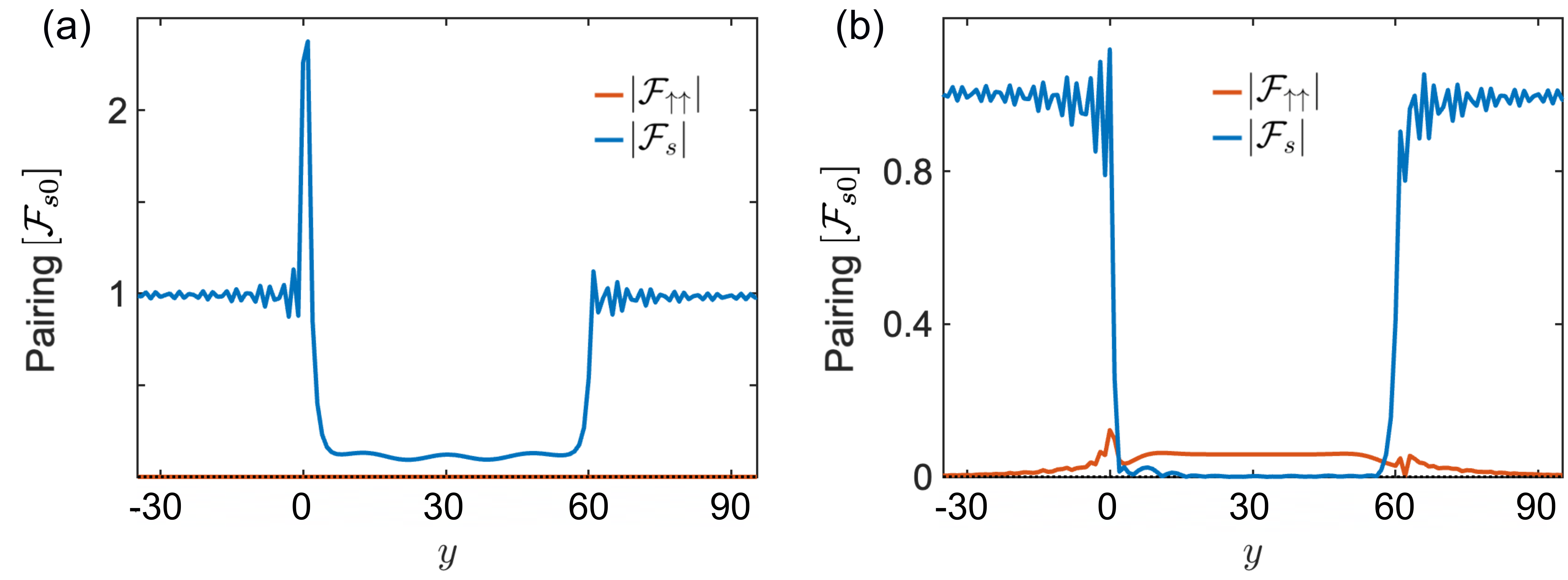}
    \caption{(a) Singlet $|\mathcal{F}_s|$ (blue) and equal-spin triplet  $|\mathcal{F}_{\uparrow\uparrow}|=|\mathcal{F}_{\downarrow\downarrow}|$ (orange) pairing amplitudes in the junction without cAFM order ($J=0$). $\mathcal{F}_{s0}$ is the singlet pairing amplitude in the bulk of the superconductor.  (b) Same as (a) but in the presence of cAFM ($J=0.4t$). The results are calculated using self-consistently determined superconducting order parameters. Parameters: $U=0.65t$ (yielding $\Delta=0.023t$ and $k_BT_c = 0.013t$),  $T=0.1T_c$, $N_S=N_L=60$, and other parameters are the same Figs.~2(a)-(b) of the main text.}
    \label{fig:F_y_self}
\end{figure}

The self-consistent calculation directly provides the singlet pairing correlations ($f_j= \langle {c_{j,\nu,\uparrow} c_{j,\nu,\downarrow}} \rangle$) at zero frequency. However, to obtain all possible pairing channels at finite frequencies $\omega_n$, we calculate the Green function, $G=(i\omega_n-H)^{-1}$, taking into account finite $\omega_n$ and incorporating the self-consistently obtained order parameter $\Delta_j$ into the junction Hamiltonian $H$. 
Figure~\ref{fig:F_y_self} presents the net pairing amplitudes $\mathcal{F}_{\alpha}=\int d{k_x}\tilde{f}_{\alpha}(k_x)$ (with $\alpha\in\{s,\uparrow\uparrow,\downarrow\downarrow,z\}$) along the junction in absence ($J=0$) and presence ($J=0.4t$) of cAFM order. For consistency and comparison, all other parameters are chosen to be the same as those in Figs.~\ref{fig:PairingCorrelation}(a) and \ref{fig:PairingCorrelation}(b).  
We find that for $\mu_{\text{AFM}}<J$, pronounced equal-spin triplet pairing dominates across the junction, while the singlet pairing is strongly suppressed. All these features remain largely unchanged compared to Fig.~\ref{fig:PairingCorrelation}. 

\begin{figure}[t]
    \centering
    \includegraphics[width=1\linewidth]{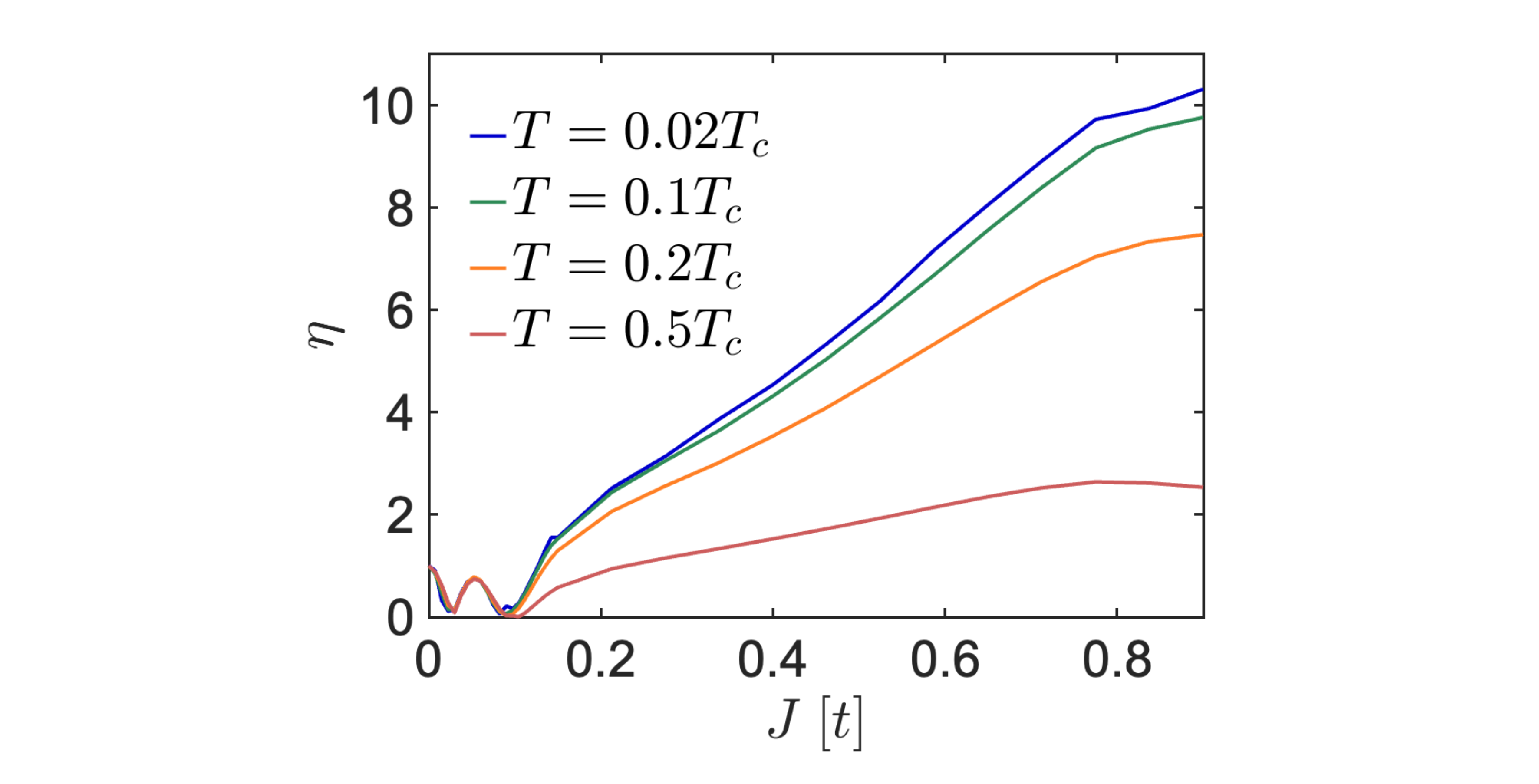}
    \caption{Enhancement ratio $\eta=I_c(J)/I_c(J=0)$ as a function of $J$ for $N_L=50$, $\mu_{\text{AFM}}=0.1t$, $\mu_S=2t$, $U=0.65t$ (yielding $\Delta=0.023t$ and $T_ck_B=0.013t$) and increasing temperature $T$.}
    \label{fig:re_IsvsJ_T}
\end{figure}

Finally, we calculate the maximum supercurrent $I_c$ as a function of $J$ at different temperatures, and other parameters are same as Fig.~\ref{fig:enhance}. Again, we find that while $I_c$ exhibit oscillations in the regime $|J|<\mu_{\text{AFM}}$, it increases overall with increasing $J$ in the regime $|J|>\mu_{\text{AFM}}$, as shown in Fig.~\ref{fig:re_IsvsJ_T}. At low temperatures, we find that $I_c$ can be enhanced by one order of magnitude compared to the non-magnetic case, and this enhancement persists over a wide range of temperatures. The supercurrent enhancement and robust $\pi$-junction states caused by the cAFM order, based on the self-consistent solution of superconducting order parameters, lead to the same main conclusions.

\section{Discussion}\label{sec:discussion}

Magnetic order is generally detrimental to superconductivity, often suppressing the supercurrent. In contrast, we demonstrate cAFM can enhance the Josephson supercurrent significantly. This enhancement arises from the generation of dominant equal-spin triplet pairing and pronounced momentum-space fluctuations of singlet pairing, both driven by cAFM. 
Furthermore, the junction undergoes a $0-\pi$ transition as a function of the cAFM strength $J$ when $|J|<|\mu_{\text{AFM}}|$ and exhibits a stable $\pi$-junction behavior for $|J|>|\mu_{\text{AFM}}|$. These effects are sustained over a broad energy window and remain stable for different temperatures and junction lengths.

cAFMs with large spin-split bands have been found in many materials~\cite{chen2014anomalous,kubler2014non,manchon2019current,Jiang24PX,Ren24PRX,xiao24PRX}.
In particular, Mn$_3$Ge, Mn$_3$Ga, and Mn$_3$Sn feature one Fermi surface at each valley over broad energy ranges around $E = -0.15$, $0.1$, and $-0.1$ eV, respectively~\cite{zhang2017strong,Qin22AM},
making them ideal candidates for testing our predictions.
Interestingly, recent experiments have observed unexpectedly large Josephson currents in junctions based on Mn$_3$Ge thin films, despite the strong Fermi-surface spin splitting~\cite{jeon2021long,jeon2023chiral}. Our magnetic enhancement via triplet pairing and singlet fluctuations offers a plausible explanation for this observation. 

\begin{acknowledgments}
We thank Lunhui Hu, Chuan Li, Jian Li, Qian Niu, Ziqiang Wang, and Zhenyu Zhang for valuable discussions. 
J.X.H. and S.B.Z. were supported by the start-up fund at HFNL, the Innovation Program for Quantum Science and Technology (Grant No. 2021ZD0302801), and the National Natural
Science Foundation of China (Grants No. 12488101)
H.P.S and B.T. were supported by the W\"urzburg-Dresden Cluster of Excellence ct.qmat (Project-id 390858490), DFG (SFB 1170), and Bavarian Ministry of Economic Affairs, Regional Development and Energy for ﬁnancial support within the High-Tech Agenda Project “Bausteine f\"ur das Quanten Computing auf Basis topologischer Materialen”.
\end{acknowledgments}

\appendix

\section{Band structures \label{sec:bandstructure}}
In absence of superconductivity and magnetism, the Hamiltonian for the kagome system can be written as
\begin{equation}
    H_0 (k_x)=  \sum_j \big\{\Psi_{j}^\dag \mathcal{H}(k_x) \Psi_{j} + \big[\Psi_{j}^\dag \mathcal{T}(k_x) \Psi_{j+1}+\mathrm{h.c.}\big]\big\},
    \label{eq:kagome-model}
\end{equation}
where $\Psi_{j} = (c_{j,1,\uparrow}, c_{j,2,\uparrow}, c_{j,3,\uparrow},c_{j,1,\downarrow}, c_{j,2,\downarrow}, c_{j,3,\downarrow})^T$, and 
$c_{j,\nu,s}$ ($c_{j,\nu,s}^\dag$) is the annihilation (creation) operator of an electron with spin $s\in\{\uparrow,\downarrow\}$ at the $\nu$-th sub-lattice of the $j$-th layer {  of unit cells along $y$-direction}.
The local Hamiltonian $\mathcal{H}(k_x)$ for each layer and the hopping matrix $\mathcal{T}(k_x)$ between neighboring layers are given by $\mathcal{H}(k_x) = t s_0 \otimes h(k_x)$ and $\mathcal{T}(k_x)= t s_0 \otimes T^+(k_x)$ with
\begin{align}
    h(k_x) = - \left(\begin{matrix}
        0 & 1+ e^{ik_x} & 1 \\
        1+ e^{-ik_x} & 0 & 1 \\
        1 & 1 & 0 \\
    \end{matrix}\right)+\left(1-\dfrac{\mu}{t}\right)\;I_{3\times3}, 
\end{align}
and 
\begin{align}
    T^+(k_x)  = - \left(\begin{matrix}
        0 & 0 & e^{ik_x} \\
        0 & 0 & 1 \\
        0 & 0 & 0 \\
    \end{matrix}\right). 
    \label{eq:Tmatrix}
\end{align}
Here, $s_0$ denotes the unit matrix in spin space, $I_{3\times3}$ is the three-by-three identity matrix, and $t$ is the hopping amplitude within and between unit cells. 
In full momentum space, the Hamiltonian is given by
\begin{equation} \label{eq-ham0-kagome}
{\cal H}_{0}({\bf k)}  = - 2t s_0 \otimes \begin{pmatrix}- 1/2& \cos({\bf k}\cdot{\bf a}_{1}) & \cos({\bf k}\cdot{\bf a}_{2})\\
\cos({\bf k}\cdot{\bf a}_{1}) & -1/2 & \cos({\bf k}\cdot{\bf a}_{3})\\
\cos({\bf k}\cdot{\bf a}_{2}) & \cos({\bf k}\cdot{\bf a}_{3}) & -1/2
\end{pmatrix},
\end{equation}
where ${\bf a}_{1}=(1,0)$, ${\bf a}_{2}=(1,\sqrt{3})/2$, and ${\bf a}_{3}={\bf a}_{2}-{\bf a}_{1}$ denote the three nearest-neighbor vectors of the kagome lattice. The kagome model processes two typical types of electrons~\cite{SBZhang25arXiv-b}. The first type with energies close to the band edge at the $\Gamma$ point behaves like Schr\"odinger electrons with quadratic energy dispersion. The second type, located near the band center ($E=0$) at the $K/K'$ points of the hexagonal Brillouin zone, behaves like Dirac fermions with linear dispersion. In absence of magnetic order, the bands are spin degenerate, forced by time-reversal symmetry. 

\begin{figure*}[t]
\centering
\includegraphics[width=0.98\linewidth]{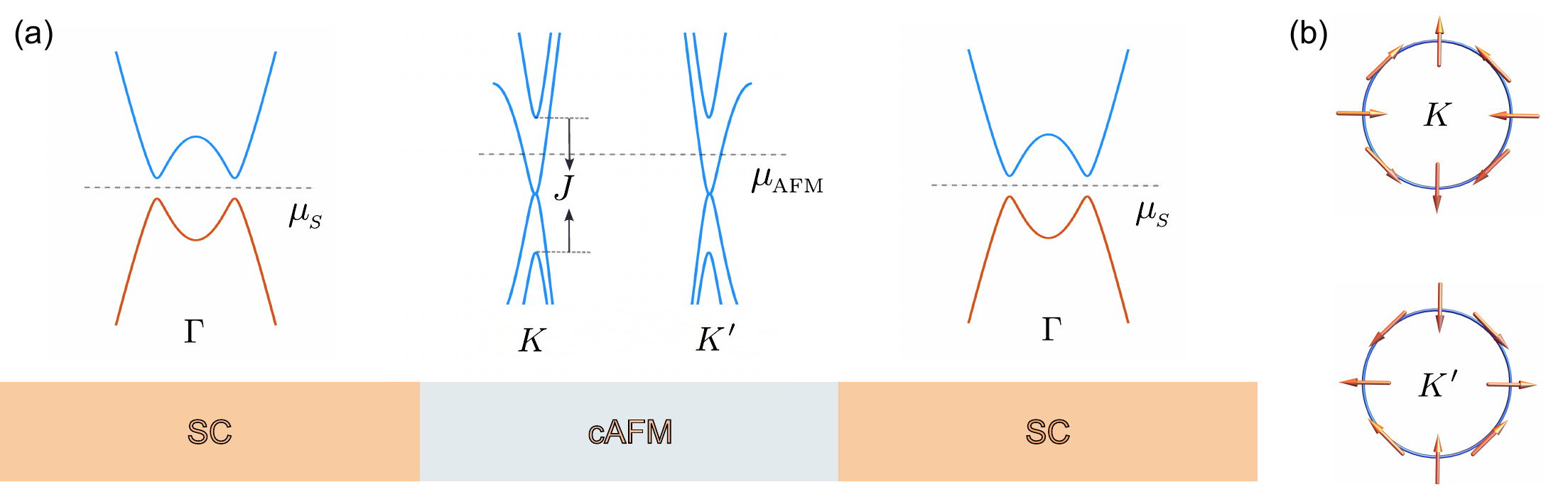}
\caption{(a) Band structures of the superconductors and cAFM.  The superconducting leads have a large { Fermi energy}, and its normal state (without pairing potential) has a parabolic band structure, resembling conventional electron gases. The pairing potential opens a full pairing gap at the Fermi surface. The Bogoliubov band structure is displayed in the panel, where the blue and orange curves represent the electron-like and hole-like bands, respectively. The cAFM has the { Fermi energy} near the band center, where the bands are strongly split by the cAFM order. Thus, in a broad range of energy, there is only one Fermi surface at each valley.  
(b) Spin textures (represented by arrows) on the Fermi surfaces (blue circles) located at the $K$ and $K^\prime$ points.
}
\label{fig:band-structure}
\end{figure*}

The band structures for each region of the junction are sketched in Fig.~\ref{fig:band-structure}(a).
In the superconductors, we consider a { large Fermi energy} such that a large Fermi surface appears at the $\Gamma$ point, as typically seen in conventional Bardeen-Cooper-Schrieffer (BCS) superconductors. The pairing potential $\Delta$ opens a band gap of magnitude $2\Delta$ in the quasiparticle energy spectrum. In the cAFM region, the Fermi energy is chosen near the band centre which features Dirac electrons at the $K$ and $K'$ points. The local non-collinear magnetic moment in the cAFM breaks the degeneracy of the Dirac cones at the $K/K^{\prime}$ points, with the energy splitting determined by the magnetic order strength $J$. 
For small { Fermi energies} $|\mu_{\text{AFM}}|$ ($<|J|$, measured from the Dirac point), one Fermi surface appears at each valley (i.e., the $K$ and $K'$ points). 
The two Fermi surfaces at the two valleys exhibit opposite in-plane spin textures~\cite{SBZhang25arXiv-a}, as shown in Fig.~\ref{fig:band-structure}(b). The spin texture is of even-parity, i.e., ${\bf S}({\bf k})={\bf S}(-{\bf k})$. For $|\mu_{\text{AFM}}|>|J|$, two Fermi surfaces appear at each valley. 



\section{Comparisons of results from two methods}\label{sec:compare_two _methods}

The Josephson supercurrent $I_s$ across the junction can be obtained from the free energy $F(\phi)$ of the system, using the formula~\cite{Tinkham2004Book}: 
\begin{equation}
    I_s = \frac{2e}{\hbar}\frac{\partial F(\phi)}{\partial\phi},
\end{equation}
where $\phi$ is the superconducting phase difference between the superconducting leads. 
The free energy $F$ at temperature $T$ is determined by the partition function $Z=\sum_{E_i}e^{- E_i/(k_BT)}$ through the relation $F=- k_B T\ln{Z}$. Here, $E_i$ is the total energy of the system in the respective microstate. At zero temperature, the free energy corresponds to the ground state energy, calculated as the sum of all negative eigenenergies of the single-particle Hamiltonian. The eigenenergies are obtained by exactly diagonalizing the Hamiltonian.  
In the calculation, we consider a long length ($N_S=100$ layers) for the superconducting regions and a temperature-dependent pairing potential described by~\cite{Tinkham2004Book} 
\begin{align}
\label{eq:T-dependence}
    \Tilde{\Delta}(T) = \Delta \tanh{\left(\frac{\pi k_B T_c}{\Delta}\sqrt{\frac{T_c}{T}-1}\right)},
\end{align}
where $T_c=0.57\Delta/k_B$ is the critical temperature. 

Alternatively, the Josephson current can be calculated using the standard Green function method~\cite{Asano01prb,SBZhang20PRB}.  
In the Matsubara formalism, the Josephson current can be found as
\begin{align}
I_s = & - \dfrac{ie}{2\hbar} k_BT \int\dfrac{dk_x}{2\pi} \sum_{\omega_n} \text{Tr} [\check{\tau}_3 \check{T}_+ \check{G}(j,j+1)  \notag \\ 
& \;\;\;\; -\check{\tau}_3 \check{T}_- \check{G}(j+1,j)],
\label{eq:current}
\end{align} 
where $\omega_n=(2n + 1)\pi k_BT$ is the Matsubara frequency for fermions with $n$, $k_B$ and $T$ being integer numbers, Boltzmann constant and temperature, respectively. In the above formulas, $\check{\cdots}$ indicates a $12\times 12$ matrix; $\check{T}_+ = \check{T}_-^\dagger$ is the hopping matrix between the neighbouring unit cells along $y$ direction; $\check{\tau}_3$ is the third Pauli matrix in particle-hole space; and Tr$[\cdots]$ means tracing over spin, particle-hole and sub-lattice spaces. We omit the dependence on $k_x$ and $\omega_n$ in the Green functions and the $T_\pm$ functions, i.e., $\check{G}(j,j'):= \check{G}_{\omega_n}(j,j';k_x)$ and $\check{T}_\pm := \check{T}_\pm(k_x)$. The Green functions are given by
\begin{subequations}
\begin{align}
\check{G}_{} (j+1,j) = & \; \check{G}_{}({j+1},j+1)
 \check{T}_+ \check{G}_{}(j,j), \\
 \check{G}_{}(j,j+1) = & \; 
 \check{G}_{}(j,j) \check{T}_- \check{G}_{}({j+1},j+1).
\end{align}
\label{eq:recGF}
\end{subequations}
We solve $\check{G}_{}({j},j)$ numerically using the recursive Green function approach~\cite{Sancho1984JPFMP,Sancho1985JPFMP}. In the numerical calculation, it is convenient to compute the current inside the cAFM region, where charge is conserved. Due to current continuity, the current is the same at any site within the region [Fig.~\ref{fig:Js_y_and_Js_phi}]. Notably, in this approach, the Green function of the entire system is iteratively constructed by adding slices of the system at a time, allowing the calculation of physical quantities, such as the current, in a computationally efficient manner.

\begin{figure}[t]
\centering
\includegraphics[width=1\linewidth]{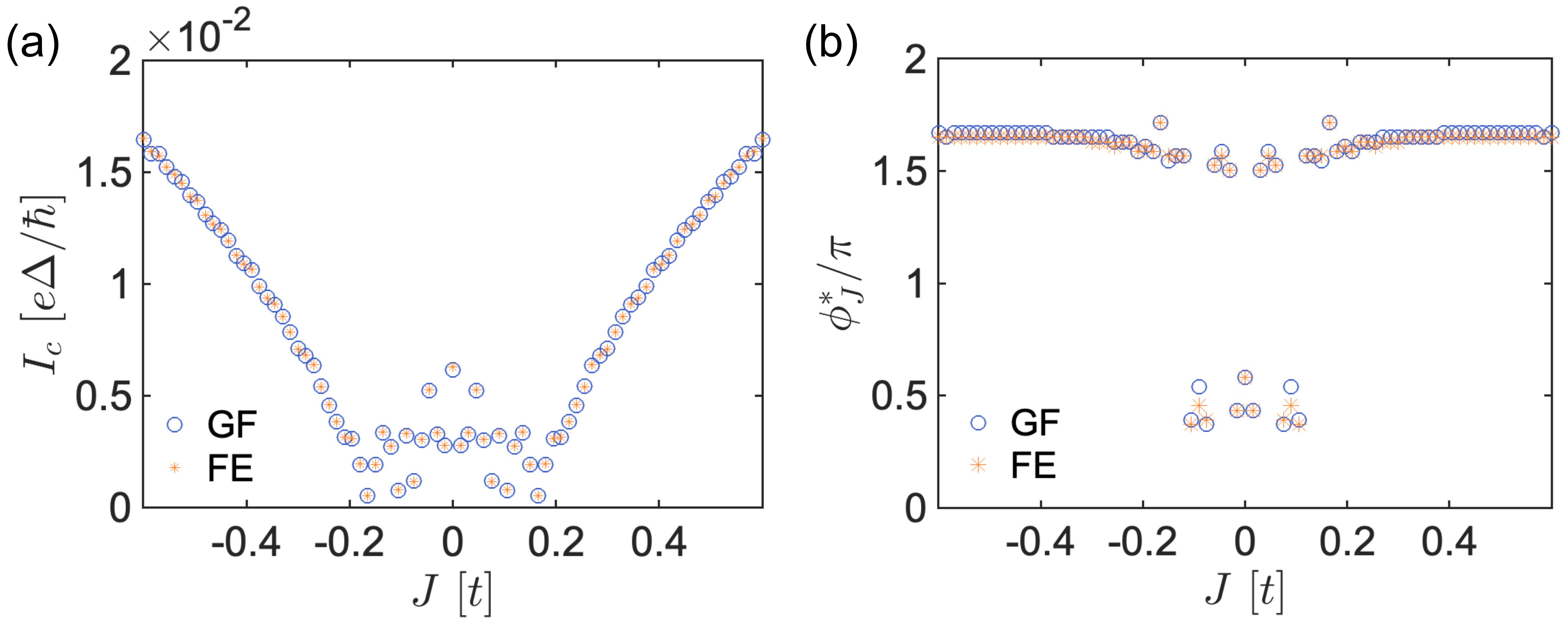}
\caption{(a) Maximum current $I_c$ and (b) its phase position $\phi_I^*$ as functions of cAFM strength $J$. The blue circles and red stars present the results calculated using the recursive Green function and free energy methods respectively. 
In the Green function method, we use a low temperature $k_B T=0.02\Delta$. In the free-energy method, we consider zero temperature ($T=0$) and set the length of the superconducting regions to $N_S=100$ (in units of the unit cell). The results from the two methods agree excellently with each other. Other parameters are $\mu_{\text{AFM}}=0.2t$, $\mu_S=2t$, $\Delta=0.02t$, and $N_L=60$.
}
\label{fig:compa_GF_FE_Ic_vs_J}
\end{figure}

In Figs.~\ref{fig:compa_GF_FE_Ic_vs_J}(a) and \ref{fig:compa_GF_FE_Ic_vs_J}(b), we compare the maximum supercurrent $I_c$ and its corresponding phase position $\phi^*_I$ obtained via the two different methods. For this illustration, we consider zero temperature ($T = 0$) for the free-energy method, while a finite low temperature ($k_BT=0.02\Delta$) for the Green function method. Other parameters are $\mu_{\text{AFM}}=0.2t$, $\mu_S=2t$, $\Delta=0.02t$, and $N_L=60$.  
The results from these two methods are consistent, confirming their accuracy.

\section{Analysis of Josephson supercurrents \label{Sec:Analysis}}
To better understand the connection between the supercurrent and pairing correlations and to gain insight into the supercurrent enhancement, we analyze the formula of the supercurrent in Eq.~\eqref{eq:current}, focusing on the cAFM region. 
In Nambu space, the Green function can be written in the form
\begin{align}
\check{G}({j,j}) &= \; \begin{pmatrix}
        \hat{G}_{e,j} & \hat{F}_{eh,j} \\
        \hat{F}_{he,j} & \hat{G}_{h,j}
    \end{pmatrix}, \;\;\;\;\; 
\label{eq:Gfunction}
\end{align}
where $\hat{G}_{e(h),j}$ is the Green function of the electrons (holes) at layer $j$, and $\hat{F}_{eh(he),j}$ is the anomalous Green function. 
The hopping matrices in particle-hole basis can be expressed as
\begin{align}
\check{T}_+ = \begin{pmatrix}
        \hat{T}_{e}^+ & \hat{0} \\
        \hat{0} & \hat{T}_{h}^+
    \end{pmatrix}, \;
\check{T}_- = \begin{pmatrix}
        \hat{T}_{e}^- & \hat{0} \\
        \hat{0} & \hat{T}_{h}^-
    \end{pmatrix}. \label{eq:HopMat_pm}
\end{align}
Plugging Eqs.~\eqref{eq:Gfunction} and \eqref{eq:HopMat_pm} into Eqs.~\eqref{eq:current} and \eqref{eq:recGF}, we find 
\begin{align}
\notag
I_s 
= & -\dfrac{ie}{2\pi\hbar} k_BT \int {dk_x} 
  \sum_{\omega_n} \text{Tr} \big[ \hat{T}_{e}^+ \hat{F}_{eh,j} 
    \hat{T}_{h}^- \hat{F}_{he,j+1} \\
  & \;\;\;\;\;
  - \hat{T}_{h}^+ \hat{F}_{he,j} \hat{T}_{e}^- \hat{F}_{eh,j+1} \big]. 
 \label{eq:kernel_fun}
\end{align}
This result tells that the supercurrent is essentially determined by a convolution of two anomalous Green functions $\hat{F}$ (i.e., pairing correlations). 

In spin space, we further decompose the anomalous Green functions and hopping matrices as 
\begin{subequations}
\begin{align}
\hat{F}_{eh,j} = &\; 
\begin{pmatrix}
    \tilde{f}_{\uparrow\uparrow,j} & \tilde{f}_{\uparrow\downarrow,j} \\
    \tilde{f}_{\downarrow\uparrow,j} & 
    \tilde{f}_{\downarrow\downarrow,j}
\end{pmatrix}, \;\;\;\;\; \\
\hat{F}_{he,j} = &\; 
\begin{pmatrix}
    \tilde{f}'_{\uparrow\uparrow,j} & 
    \tilde{f}'_{\uparrow\downarrow,j} \\
    \tilde{f}'_{\downarrow\uparrow,j} & 
    \tilde{f}'_{\downarrow\downarrow,j}
\end{pmatrix}, \;\;\;\;\; \\
    \hat{T}_{e(h)}^{+(-)} = &\; 
\begin{pmatrix}
    T_{e(h)\uparrow}^{+(-)} & 0 \\
    0 & T_{e(h)\downarrow}^{+(-)}
\end{pmatrix}.
\end{align}
\end{subequations}
Note that in our system, the hopping matrices are diagonal in spin space. $\tilde{\cdots}$ indicates a three-by-three matrix in sublattice space. 
Plugging these decompositions into the trace in Eq.~\eqref{eq:kernel_fun}, we obtain

\begin{align}
 & \; \text{Tr}\big[ 
        \hat{T}_{e}^+ \hat{F}_{eh,j} \hat{T}_{h}^- \hat{F}_{he,j+1} 
       -\hat{T}_{h}^+ \hat{F}_{he,j} \hat{T}_{e}^- \hat{F}_{eh,j+1} 
    \big]   \notag \\
= & \; \text{Tr}\big[ 
     T_{e\uparrow}^{+} \tilde{f}_{\uparrow\downarrow,j} T_{h\downarrow}^{-} \tilde{f}'_{\downarrow\uparrow,j+1} 
     - T_{h\uparrow}^{+} \tilde{f}'_{\uparrow\downarrow,j}  T_{e\downarrow}^{-}\tilde{f}_{\downarrow\uparrow,j+1} \big] \notag \\ 
    & + \text{Tr}\big[T_{e\downarrow}^{+}\tilde{f}_{\downarrow\uparrow,j} T_{h\uparrow}^- \tilde{f}'_{\uparrow\downarrow,j+1} 
    - T_{h\downarrow}^{+} \tilde{f}'_{\downarrow\uparrow,j} T_{e\uparrow}^-\tilde{f}_{\uparrow\downarrow,j+1} 
  \big] \notag \\  
  & + \text{Tr}\big[ 
    T_{e\uparrow}^{+}\tilde{f}_{\uparrow\uparrow,j} T_{h\uparrow}^- \tilde{f}'_{\uparrow\uparrow,j+1} 
  - T_{h\uparrow}^{+} \tilde{f}'_{\uparrow\uparrow,j} T_{e\uparrow}^-\tilde{f}_{\uparrow\uparrow,j+1} 
  \big] \notag \\
  & + \text{Tr}\big[ 
     T_{e\downarrow}^{+} \tilde{f}_{\downarrow\downarrow,j} T_{h\downarrow}^{-} \tilde{f}'_{\downarrow\downarrow,j+1}
    - T_{h\downarrow}^{+} \tilde{f}'_{\downarrow\downarrow,j} T_{e\downarrow}^{-} \tilde{f}_{\downarrow\downarrow,j+1}
  \big].
\label{eq:kernel_fun-2}
\end{align}
This indicates that the opposite-spin and two equal-spin pairing correlations contribute separately to $I_s$, i.e.,
\begin{equation}I_s=I_{s\uparrow\downarrow}+I_{s\uparrow\uparrow}+I_{s\downarrow\downarrow},
\end{equation} where
\begin{subequations}
\begin{align}
I_{s\uparrow\uparrow} = &- \dfrac{ie}{2\pi\hbar} k_BT \int {dk_x}  \sum_{\omega_n} 
\text{Tr} \big[ T_{e\uparrow}^{+} \tilde{f}_{\uparrow\uparrow,j} T_{h\uparrow}^- \tilde{f}'_{\uparrow\uparrow,j+1} \notag \\
&  \;\;\;\; - T_{h\uparrow}^{+} \tilde{f}'_{\uparrow\uparrow,j} T_{e\uparrow}^-\tilde{f}_{\uparrow\uparrow,j+1} 
  \big], \\
I_{s\downarrow\downarrow} = &- \dfrac{ie}{2\pi\hbar} k_BT \int {dk_x} \sum_{\omega_n} 
\text{Tr} \big[ 
     T_{e\downarrow}^{+} \tilde{f}_{\downarrow\downarrow,j} T_{h\downarrow}^{-} \tilde{f}'_{\downarrow\downarrow,j+1} \notag  \\
& \;\;\;\;  - T_{h\downarrow}^{+} \tilde{f}'_{\downarrow\downarrow,j} T_{e\downarrow}^{-} \tilde{f}_{\downarrow\downarrow,j+1}
  \big],\\
I_{s\uparrow\downarrow} = &- \dfrac{ie}{2\pi\hbar} k_BT \int {dk_x}  \sum_{\omega_n} 
\text{Tr} \big[ 
     T_{e\uparrow}^{+} \tilde{f}_{\uparrow\downarrow,j} T_{h\downarrow}^{-} \tilde{f}'_{\downarrow\uparrow,j+1} \notag \\
& \;\;\;\; + T_{e\downarrow}^{+} \tilde{f}_{\downarrow\uparrow,j} T_{h\uparrow}^- \tilde{f}'_{\uparrow\downarrow,j+1} - T_{h\uparrow}^{+} \tilde{f}'_{\uparrow\downarrow,j}  T_{e\downarrow}^{-}\tilde{f}_{\downarrow\uparrow,j+1}  \notag \\
    & 
    \;\;\;\; - T_{h\downarrow}^{+} \tilde{f}'_{\downarrow\uparrow,j} T_{e\uparrow}^-\tilde{f}_{\uparrow\downarrow,j+1} \big].
\end{align}
\end{subequations}
Here, $\tilde{f}_{ss^\prime,j}$ and $\tilde{f}_{ss^\prime,j}^\prime$ denote the anomalous Green functions with spin indices $s$ and $s^\prime$. 

\begin{figure}[t]
\centering
\includegraphics[width=1\linewidth]{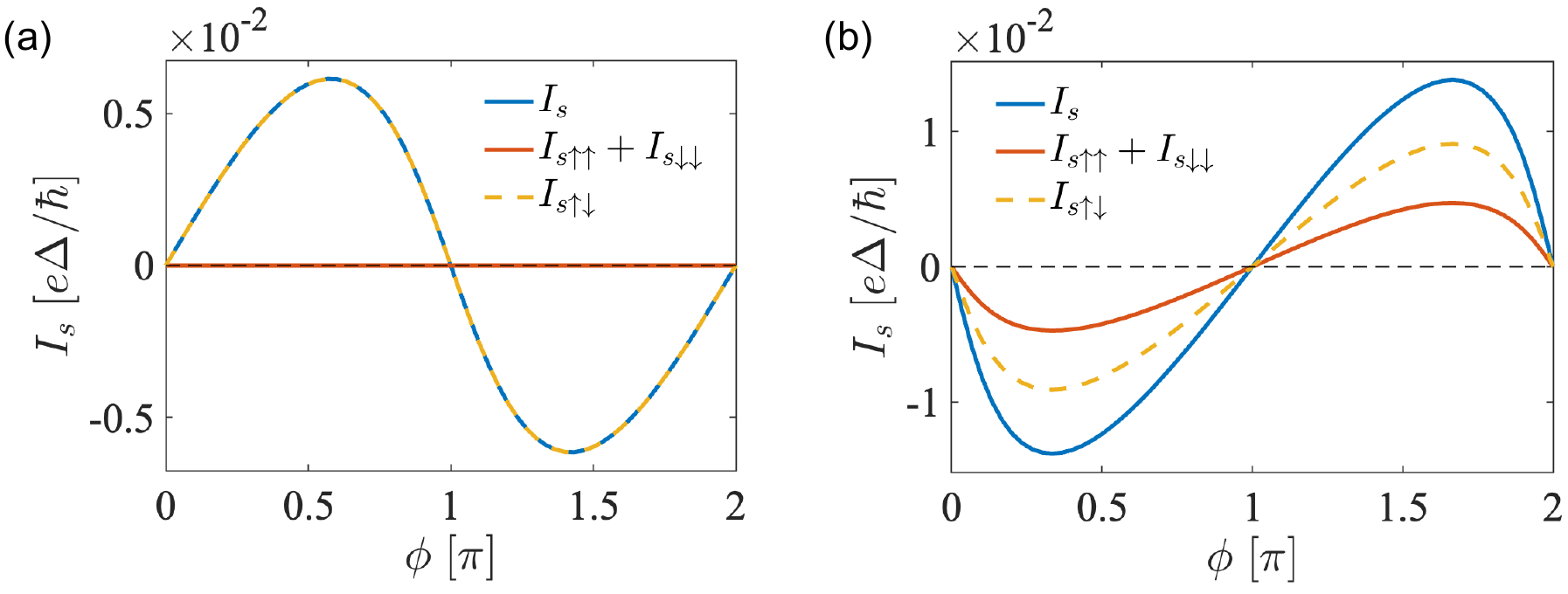} 
\caption{(a) Supercurrent contributions from equal-spin-triplet pairing $I_{s\uparrow\uparrow}+I_{s\downarrow\downarrow}$ (orange) and spin-singlet pairing $I_{s\uparrow\downarrow}$ (yellow) and total supercurrent $I_s=I_{s\uparrow\downarrow}+I_{s\uparrow\uparrow}+I_{s\downarrow\downarrow}$ (blue) as functions of $\phi$ in the absence of cAFM ($J=0$). (b) Same as (a) but for the presence of cAFM ($J=0.5t$).
Other parameters: $\Delta=0.02t$, $\mu_{\text{AFM}}=0.2t$, $\mu_S=2t$ and $N_L=60$. }
\label{fig:Is_uu_ud_vs_phi}
\end{figure}

\begin{figure*}[t]
\centering
\includegraphics[width=0.98\linewidth]{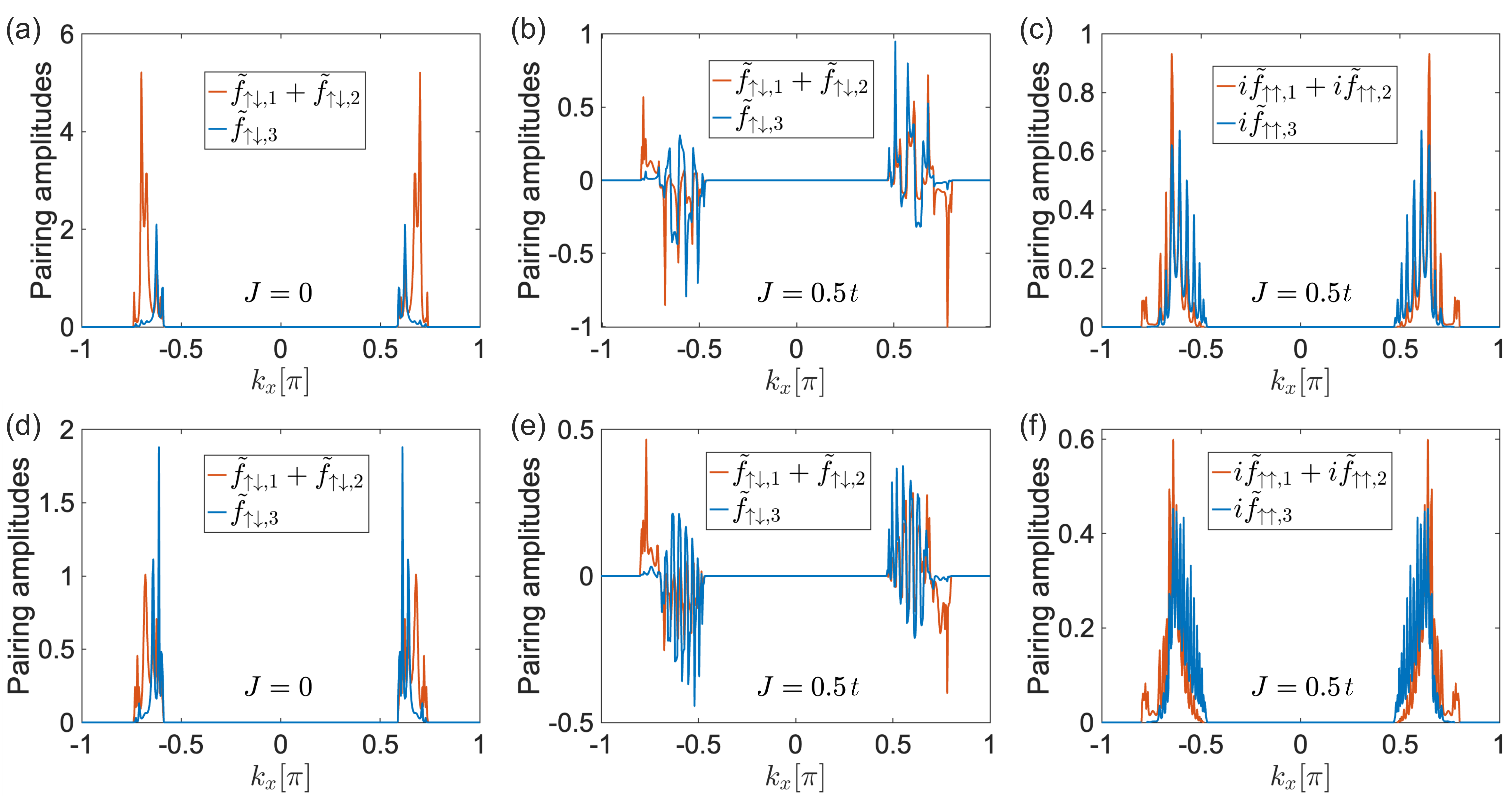} 
\caption{(a) Singlet pairing amplitudes, $\tilde{f}_{\downarrow\uparrow,1}+\tilde{f}_{\downarrow\uparrow,2}$ and $\tilde{f}_{\downarrow\uparrow,3}$, at the junction center ($y=N_L/2$) as functions of $k_x$ in absence of cAFM order ($J=0$) for $N_L=60$. The pairing amplitudes are most pronounced near the two valleys and maintain the same sign for all $k_x$. (b) Same as (a) but for the presence of cAFM ($J=0.5t$). The pairing amplitudes are most pronounced near the two valleys and strongly fluctuate around zero. Their sum over $k_x$ is vanishingly small.  (c) Same as (b) but for the equal-spin triplet pairing amplitudes,  $\tilde{f}_{\uparrow\uparrow,1}+\tilde{f}_{\uparrow\uparrow,2}$ and $\tilde{f}_{\uparrow\uparrow,3}$. The pairing amplitudes are most pronounced near the two valleys and have the same sign for all $k_x$. (d)-(f) are the same as (a)-(c) but for a longer junction length ($N_L=90$). The fluctuations become more rapid for $N_L=90$. 
Other parameters for all panels are $\Delta=0.02t$, $\mu_S=2t$, $\mu_{\text{AFM}}=0.2t$, $\omega_n=\pi k_BT$, and $k_BT=0.02\Delta$.} 
\label{fig:fluctuation}
\end{figure*}

Typical current-phase relations, $I_{s,\uparrow\uparrow}+I_{s,\downarrow\downarrow}$ and $I_{s,\uparrow\downarrow}$, are presented in Fig.~\ref{fig:Is_uu_ud_vs_phi}. These supercurrents are calculated at the junction center ($y=N_L/2$), where the current conservation is ensured.
When the cAFM order is absent ($J=0$), only singlet pairing is allowed [Figs.~\ref{fig:Is_uu_ud_vs_phi}(b) and \ref{fig:fluctuation}(a)], leading to a conventional $0$-junction. For nonzero $J$, triplet Cooper pairs are induced [Fig.~\ref{fig:fluctuation}(c)]. For strong $J$ ($J>\mu_{\text{AFM}}$), the cAFM supports only net triplet pairing correlations. Nevertheless, despite the vanishing net singlet pairing correlation (after integrating over $k_x$, i.e., $\int dk_x \tilde{f}_{\uparrow\downarrow(\downarrow\uparrow),j,\nu}(k_x)=0$), singlet pairing still contributes significantly to the supercurrent [Fig.~\ref{fig:Is_uu_ud_vs_phi}(b)]. This stems from strong fluctuations of singlet pairing in momentum space, where for individual $k_x$, $\tilde{f}_{\downarrow\uparrow(\uparrow\downarrow),j}$ can be nonzero, as illustrated in Fig.~\ref{fig:fluctuation}(b).   

    
To get further insight into the supercurrent, we analyze the pairing correlations in sublattice space. 
For the kagome system, the hopping matrices satisfy $T^{+(-)}_{es} = -T^{+(-)}_{hs} = T^{+(-)}$, and $T^+ = (T^-)^\dag$ with
\begin{align}
    T^+  = - \left(\begin{matrix}
        0 & 0 & te^{ik_x} \\
        0 & 0 & t \\
        0 & 0 & 0 \\
    \end{matrix}\right).
    \label{eq:Tmatrix}
\end{align}
Thus, we obtain
\begin{align}
    \text{Tr} \big[T^+ \tilde{f} T^- \tilde{f}^\prime \big] &= t^2 f_{33} (f^\prime_{11} + f^\prime_{22} + e^{-i k_x} f^\prime_{12} + e^{i k_x} f^\prime_{21}), \notag \\
    \text{Tr} \big[T^+ \tilde{f}^\prime T^- \tilde{f}\big] &= t^2 f^\prime_{33} (f_{11} + f_{22} + e^{-i k_x} f_{12} + e^{i k_x} f_{21} ),
\end{align}
where $f_{\mu\nu}$ and $f^\prime_{\mu\nu}$ denote matrix elements of $\tilde{f}$ and $\tilde{f}^\prime$ with sublattice indices $\mu,\nu\in\{1,2,3\}$, respectively.
The Josephson current from the triplet pairing can then be rewritten as 
\begin{eqnarray}
 I_{s\uparrow\uparrow} + I_{s\downarrow\downarrow} 
=I_{t}^{(1)} + I_{t}^{(2)},
\end{eqnarray}
where
\begin{subequations}
   \begin{eqnarray}
    I_{t}^{(1)} &=& - \dfrac{iet^2}{2\pi\hbar} k_BT \int {dk_x}  \sum_{\omega_n}  \notag \\
    & & \big[f^\prime_{\uparrow\uparrow,j,33} (f_{\uparrow\uparrow,j+1,11} + f_{\uparrow\uparrow,j+1,22})  \\
    & & - f_{\uparrow\uparrow,j,33} (f^\prime_{\uparrow\uparrow,j+1,11} + f^\prime_{\uparrow\uparrow,j+1,22}) +(\uparrow\rightarrow\downarrow) \big], \notag \\ \notag 
    I_{t}^{(2)} &=& - \dfrac{iet^2}{2\pi\hbar} k_BT \int {dk_x}  \sum_{\omega_n} \notag \\
    & & \big[ f^\prime_{\uparrow\uparrow,j,33} (e^{-i k_x} f_{\uparrow\uparrow,j+1,12} + e^{i k_x} f_{\uparrow\uparrow,j+1,21})  \\ 
    & &  - f_{\uparrow\uparrow,j,33} (e^{-i k_x} f^\prime_{\uparrow\uparrow,j+1,12} + e^{i k_x} f^\prime_{\uparrow\uparrow,j+1,21})   +  (\uparrow\rightarrow\downarrow) \big]. \notag 
\end{eqnarray} 
\end{subequations}
Similarly, the Josephson current from the opposite-spin pairing is written as
\begin{eqnarray}
 I_{s\uparrow\downarrow}  
    &=& I_{s}^{(1)} + I_{s}^{(2)},
\end{eqnarray}
where 
\begin{subequations}
\begin{eqnarray}
    I_{s}^{(1)} &=& - \dfrac{iet^2}{2\pi\hbar} k_BT \int {dk_x}  \sum_{\omega_n} \notag \\
    & & \big[ f^\prime_{\uparrow\downarrow,j,33} ( f_{\downarrow\uparrow,j+1,11} +  f_{\downarrow\uparrow,j+1,22} )  \notag \\
    & &  - f_{\uparrow\downarrow,j,33} ( f^\prime_{\downarrow\uparrow,j+1,11} +  f^\prime_{\downarrow\uparrow,j+1,22} )  + (\uparrow\leftrightarrow\downarrow) \big] , \\ \notag
    I_{s}^{(2)} &=& - \dfrac{iet^2}{2\pi\hbar} k_BT \int {dk_x}  \sum_{\omega_n} \notag \\
    & & \big[f^\prime_{\uparrow\downarrow,j,33} (e^{-i k_x} f_{\downarrow\uparrow,j+1,12}  + e^{i k_x}  f_{\downarrow\uparrow,j+1,21} )  \\
    & & 
    - f_{\uparrow\downarrow,j,33} ( e^{-i k_x} f^\prime_{\downarrow\uparrow,j+1,12} + e^{i k_x} f^\prime_{\downarrow\uparrow,j+1,21})  + (\uparrow\leftrightarrow\downarrow) \big]. \notag 
\end{eqnarray}
\end{subequations}
While $I_{t(s)}^{(1)}$ corresponds to the part contributed solely by the sublattice-diagonal pairing correlations,  $I_{t(s)}^{(2)}$ corresponds to the part involving inter-sublattice pairing correlations. The diagonal parts contribute most importantly to the supercurrent, as shown by numerical calculations. Thus, we focus on the diagonal terms in the following.

The Josephson current appears when there is a phase difference $\phi$ between the two superconductors. Without loss of generality, we assume that the pairing phase in the left superconductor is zero, while the pairing phase in the right superconducting lead is $\phi$. In the above formulas, the left anomalous Green function $\hat{F}_{eh(he),j}$ inherits the phase from the left superconducting lead, while the right anomalous Green function $\hat{F}_{eh(he),j+1}$ from the right superconducting lead. For simplicity, we neglect the mutual influence of the pairing correlations from the two superconducting sides, which should be justified when the junction transparency is low, such as in the case of a large Fermi surface mismatch. In the weak-tunneling regime, the pairing amplitudes can be approximately decomposed into a phase-independent component and a phase-dependent component, i.e., 
\begin{subequations}
\begin{align}
    f_{\alpha,j+1} = & \; e^{i\phi}\Tilde{f}_{\alpha,j+1}, \;\;\;\;\;  f_{\alpha,j+1}' = e^{-i\phi} \Tilde{f}_{\alpha,j+1}', \\
    f_{\alpha,j} = & \; \Tilde{f}_{\alpha,j}, \;\;\;\;\;  \;\;\;\;\;  \;\;\;\;\;  \;\;  f_{\alpha,j}' =  \Tilde{f}_{\alpha,j}',
\end{align}
\end{subequations}
where $\alpha\in\{\uparrow\uparrow,\downarrow\downarrow,\uparrow\downarrow,\downarrow\uparrow\}$, and $\Tilde{f}$ indicates the part independent of $\phi$. 
Thus, the total Josephson current is given by
\begin{equation}
    I_s \approx (I_{t,c}+I_{s,c})\sin \phi,
\end{equation}
where
\begin{subequations}
   \begin{eqnarray}
   I_{t,c} &=& - \dfrac{2e}{\pi\hbar} k_BT t^2 \sum_{\omega_n} \int {dk_x} \tilde{f}^*_{\uparrow\uparrow,j,3}({-k_x})  \notag \\
   & &  \;\;\;\times [\tilde{f}_{\uparrow\uparrow,j+1,1}({k_x}) + \tilde{f}_{\uparrow\uparrow,j+1,2}({k_x})] ,  \\
    I_{s,c} &=& - \dfrac{2e}{\pi\hbar} k_BT t^2 \sum_{\omega_n} \int {dk_x} \tilde{f}^*_{\uparrow\downarrow,j,3}({-k_x}) \notag \\
    & &   \;\;\;\times [\tilde{f}_{\downarrow\uparrow,j+1,1}({k_x}) + \tilde{f}_{\downarrow\uparrow,j+1,2}({k_x})]. 
\end{eqnarray}
\end{subequations}
Here, we use the subscript ${\tilde{f}}_{ss',j,\nu} = \tilde{f}_{ss',j,\nu\nu}$ for ease of notation and the relations between the phase-independent parts due to particle-hole symmetry: $\tilde{f}^\prime({ k_x})=-\tilde{f}^*({-k_x})$.
We also make use of the fact that the integrals $\int dk_x  \tilde{f}'_{\uparrow\uparrow,j,3} (\tilde{f}_{\uparrow\uparrow,1} + \tilde{f}_{\uparrow\uparrow,2})=\int dk_x  \tilde{f}'_{\downarrow\downarrow,3} (\tilde{f}_{\downarrow\downarrow,1} + \tilde{f}_{\downarrow\downarrow,2})$ and $\int dk_x  \tilde{f}'_{\uparrow\downarrow,3} (\tilde{f}_{\downarrow\uparrow,1} + \tilde{f}_{\downarrow\uparrow,2})=\int dk_x  \tilde{f}'_{\downarrow\uparrow,3} (\tilde{f}_{\uparrow\downarrow,1} + \tilde{f}_{\uparrow\downarrow,2})$ are real-valued.

\section{Self-consistent calculations}\label{sec:AP_self_consistent}

The superconducting order parameter in the Josephson junction can be calculated self-consistently from attractive electron-electron interactions. To illustrate this, we start from the Hubbard interaction and employ the mean-field theory, following the approach of Refs.~\cite{Black-Schaffer08PRB,Awoga19PRL,Setiawan2019PRB}. 
The Hubbard interaction is given by
\begin{equation} H_{U} = - \sum_{\bf r} U_{\bf r} c_{{\bf r},\nu,\uparrow}^\dag
c_{{\bf r},\nu,\uparrow} c_{{\bf r},\nu,\downarrow}^\dag c_{{\bf r},\nu,\downarrow}.
\end{equation}
Here, $U_{\bf r}$ is the position-dependent interaction strength. It is assumed to be a finite constant $U$ in the superconductor while elsewhere zero:
\begin{eqnarray}
    U_j = \begin{cases}
U, & j\leqslant 0\; \& \; j\geqslant N_L+1 \\
0, &  \text{otherwise}
\end{cases}
\end{eqnarray}
Under mean-field approximation, we derive the pairing term as 
\begin{eqnarray} 
    H_{SC} = \sum_{\bf r} \Delta_{\bf r} c_{{\bf r},\nu,\uparrow}^\dag  c_{{\bf r},\nu,\downarrow}^\dag +\text{h.c.}, 
\end{eqnarray}
where the order parameter is determined self-consistently as
\begin{align}
    \Delta_{\bf r} = U_{\bf r} \langle c_{{\bf r},\nu,\uparrow} c_{{\bf r},\nu,\downarrow} \rangle.
\end{align}
Here, $\langle \cdots \rangle$ indicates the thermal average. 
In the junction system, it is convenient to perform a partial Fourier transform (from $x$ to $k_x$). Thus, the pairing term can be rewritten as
\begin{eqnarray} 
    H_{SC} && = \sum_{j}\int d{k_x} \Delta_{j} c_{\{j,k_x\},\nu,\uparrow}^\dag c_{\{j,-k_x\},\nu,\downarrow}^\dag +\text{h.c.} \notag \\
    && =  \sum_{j}\int d{k_x}\; \Delta_{j} c_{j,\nu,\uparrow}^\dag  c_{j,\nu,\downarrow}^\dag +\text{h.c.}, 
\end{eqnarray}
and accordingly, the order parameter becomes
\begin{align}
    \Delta_{j} 
    & = U_{j} \int d{k_x}\langle {c_{j,\nu,\uparrow} c_{j,\nu,\downarrow}} \rangle .
\end{align}
Here, we omit the $k_x$ dependence for ease of notation.

\subsection{Superconducting order parameter}

We perform the self-consistent calculation in a finite-size junction: the cAFM region has $N_L$ layers, while two superconducting leads have $N_s$ layers (which should be longer than the superconducting coherence length). 
Both the magnitude and phase of the order parameter are determined self-consistently, except the phases at two outermost parts ($5$ layers) of the junction which are fixed at $0$ and $\phi$. For concreteness, we set $N_S=60$ and $\mu_S=2t$ (same as that used in the non-self-consistent calculations).  

\begin{figure}[t]
    \centering
    \includegraphics[width=1\linewidth]{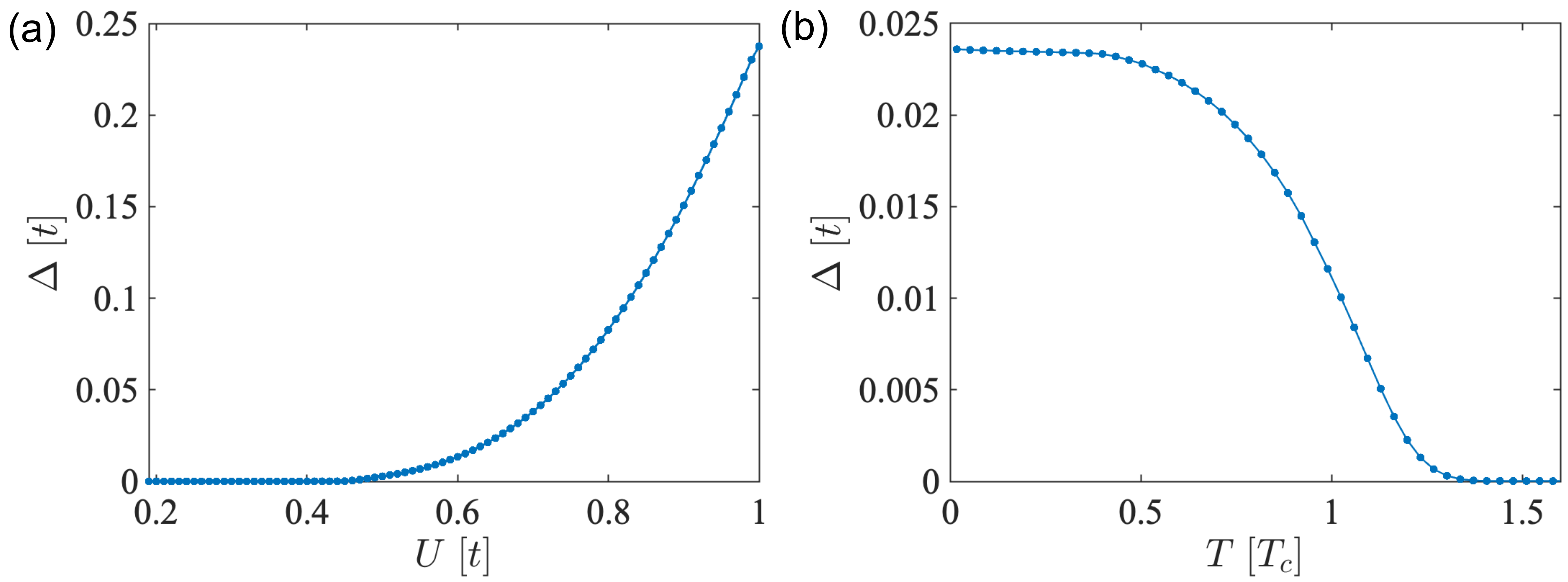}
    \caption{(a) Bulk superconducting order parameter $\Delta$ as a function of interaction strength $U$ at temperature $T=0.001t$. (b) $\Delta$ as a function of $T$ (in units of $T_c=0.57\Delta_0/k_B \approx 0.013t/k_B$, where $\Delta_0$ is the order parameter at zero temperature) for $U=0.65t$. The { Fermi energy} in the superconductors is $\mu_S=2t$.}
    \label{fig:Delta_U_T_self}
\end{figure}

Figure~\ref{fig:Delta_U_T_self}(a) shows the superconducting order parameter $\Delta$ in the bulk of the superconductor at a low temperature ($T=0.001t$) as a function of interaction strength $U$. A significant order parameter emerges for $U>0.5t$. We choose $U=0.65t$, which results in an order parameter of $\Delta\approx 0.023t$ at zero temperature. It corresponds to a superconducting coherence length of $\xi= 2t/(\pi\Delta)\approx 28$ layers, much smaller than $N_S=60$. In Fig.~\ref{fig:Delta_U_T_self}(b), we examine the temperature dependence of $\Delta$, which follows a typical BCS behavior described by Eq.~\eqref{eq:T-dependence} with critical temperature $T_c =  0.57\Delta = 0.013t/k_B$. 

\begin{figure}[t]
    \centering
    \includegraphics[width=1\linewidth]{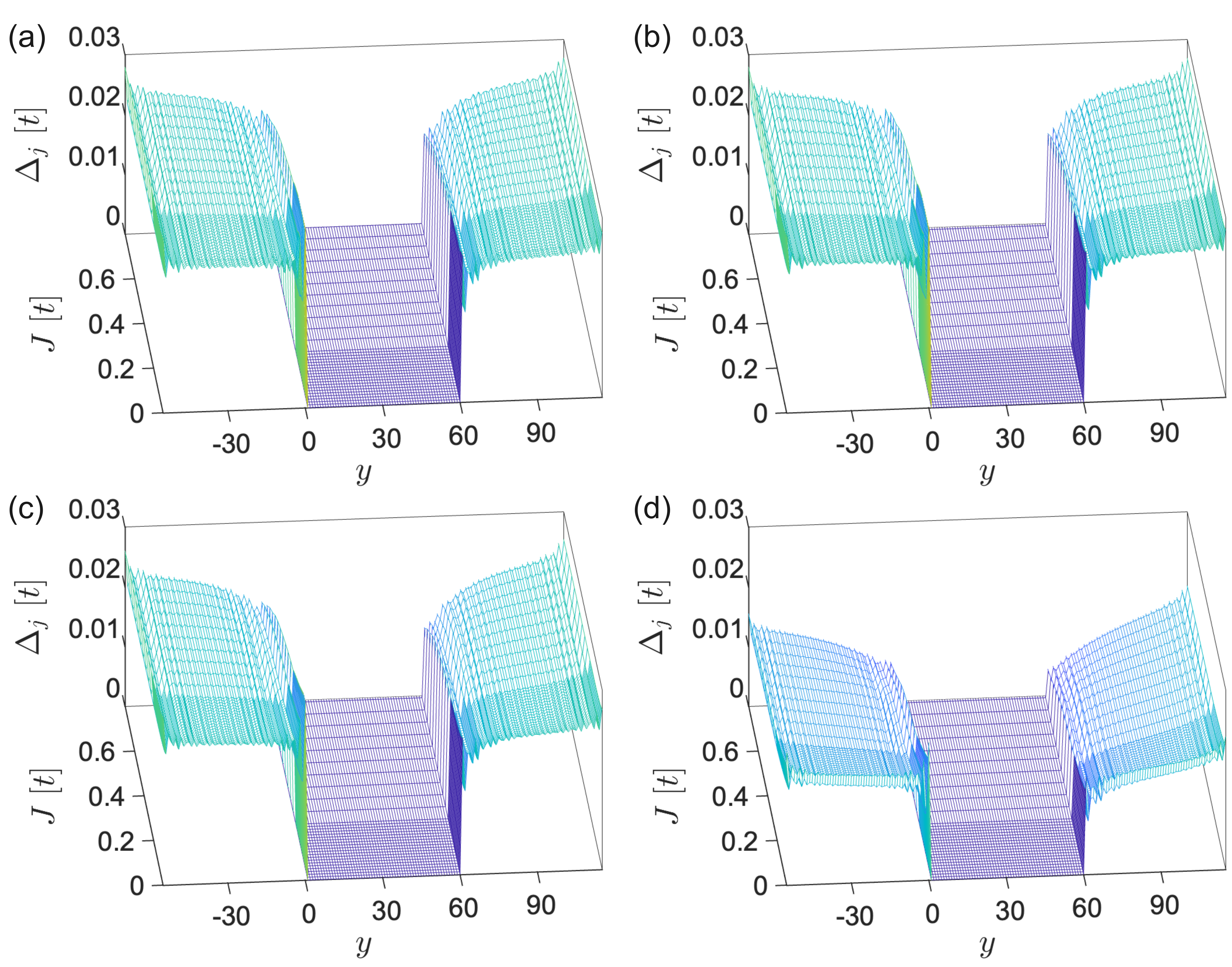}
    \caption{(a) Order parameter $\Delta_j$ in the Josephson junction for increasing cAFM strengths $J$ at temperature $T=0.1T_c$. (b)-(d) Same as (a) but for $T=0.2T_c$, $0.5T_c$ and $0.8T_c$, respectively. The results are obtained in the static limit ($\omega=0$) where only singlet pairing exists. Other parameters are $\mu_{\text{AFM}}=0.2t$, $\mu_S=2t$, $U=0.65t$ (yielding $T_c=0.013t/k_B$), and $N_L=N_S=60$. } 
    \label{fig:Delta_J_y_self}
\end{figure}

\begin{figure}[h]
    \centering
    \includegraphics[width=1\linewidth]{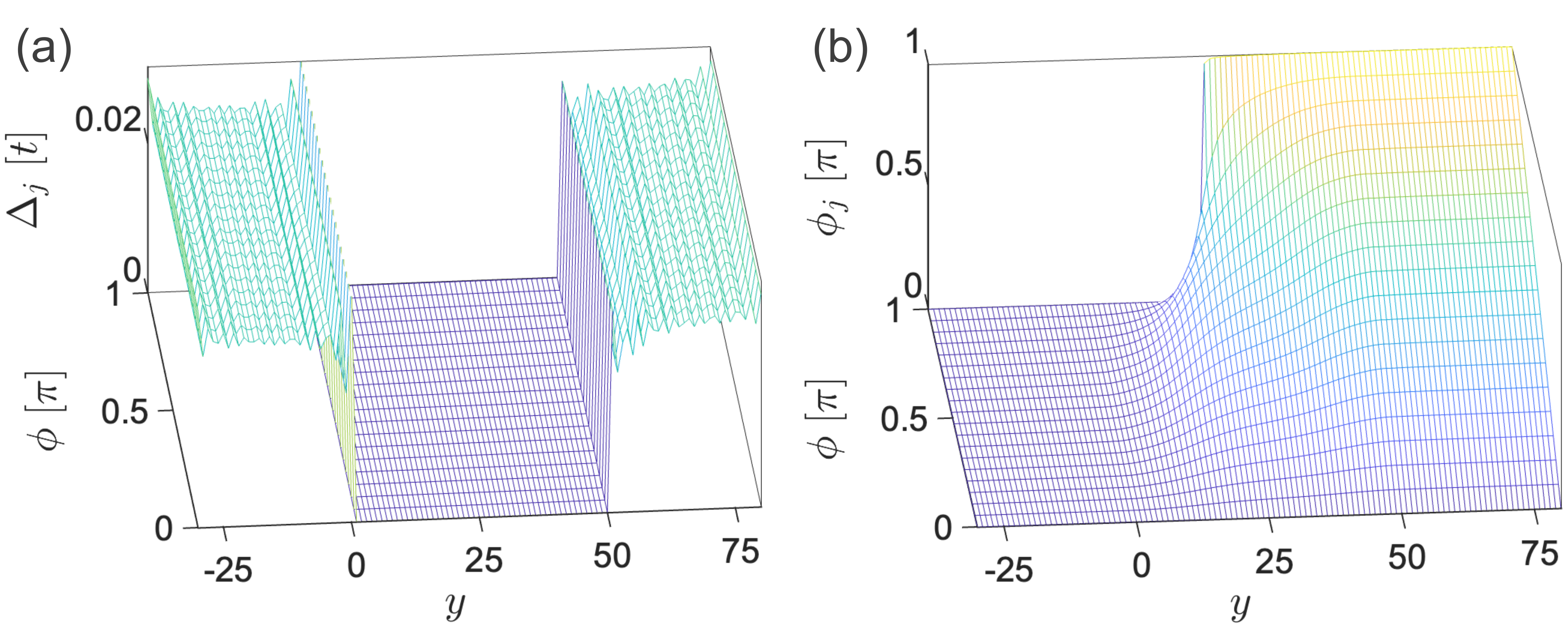}
    \caption{(a) Magnitude $|\Delta_j|$ of the local order parameter in the Josephson junction for various phase differences $\phi$, with parameters  $\mu_{\text{AFM}}=0.1t$, $\mu_S=2t$, $U=0.65t$ (yielding $T_c=0.013t/k_B$), $N_L=50$, $T=0.1T_c$ and $J=0$. (b) Phase $\phi_j$ of the local order parameter (more precisely, the pairing amplitude) corresponding to (a).} 
    \label{fig:Delta_y_phi}
\end{figure}

Next, we calculate the local superconducting order parameter $\Delta_j$ in the Josephson junction as a function of cAFM strength $J$, with parameters $\mu_S=2t$, $U=0.65t$, $\mu_{\text{AFM}}=0.2t$, $N_L=60$, $\phi=0$, and different temperatures $T$ [Fig.~\ref{fig:Delta_J_y_self}]. 
The magnitude of $\Delta_j$ decreases overall as $T$ grows, consistent with Fig.~\ref{fig:Delta_U_T_self}(b).
Notably, due to the large Fermi-surface mismatch ($\mu_S=2t$ and $\mu_{\text{AFM}}=0.2t$), the order parameter exhibits an approximately step-like profile across the interfaces, approaching its bulk value in the superconducting leads ($\Delta_j\approx \Delta$), while vanishing in the cAFM region. This feature is more pronounced for lower temperatures. 

We also compute the order parameter in the presence of phase difference self-consistently, in a similar spirit of Refs.~\cite{Black-Schaffer08PRB,Awoga19PRL}. 
As noted above, we fix the phases of the outermost five layers at the two ends of the junction to be $0$ and $\phi$, respectively. Elsewhere, the phases are determined self-consistently.  
For illustration, we present in Fig.~\ref{fig:Delta_y_phi} the magnitude and phase of the order parameter along the junction for $J=0$ and $T=0.1T_c$. The magnitude $|\Delta_j|$ shows a similar profile as in the $\phi=0$ case, remaining near constant in the superconducting leads while vanishing in the cAFM region. The local pairing phase $\phi_j$ is also approximately constant in the superconducting leads, taking values of $0$ and $\phi$, and varies smoothly from $0$ to $\phi$ across the junction [Fig.~\ref{fig:Delta_y_phi}(b)]. These results justify our assumption of constant superconducting order parameters in the superconducting leads for the case with large Fermi surface mismatch.  

\subsection{Josephson supercurrents}
\begin{figure}[t]
    \centering
    \includegraphics[width=0.98\linewidth]{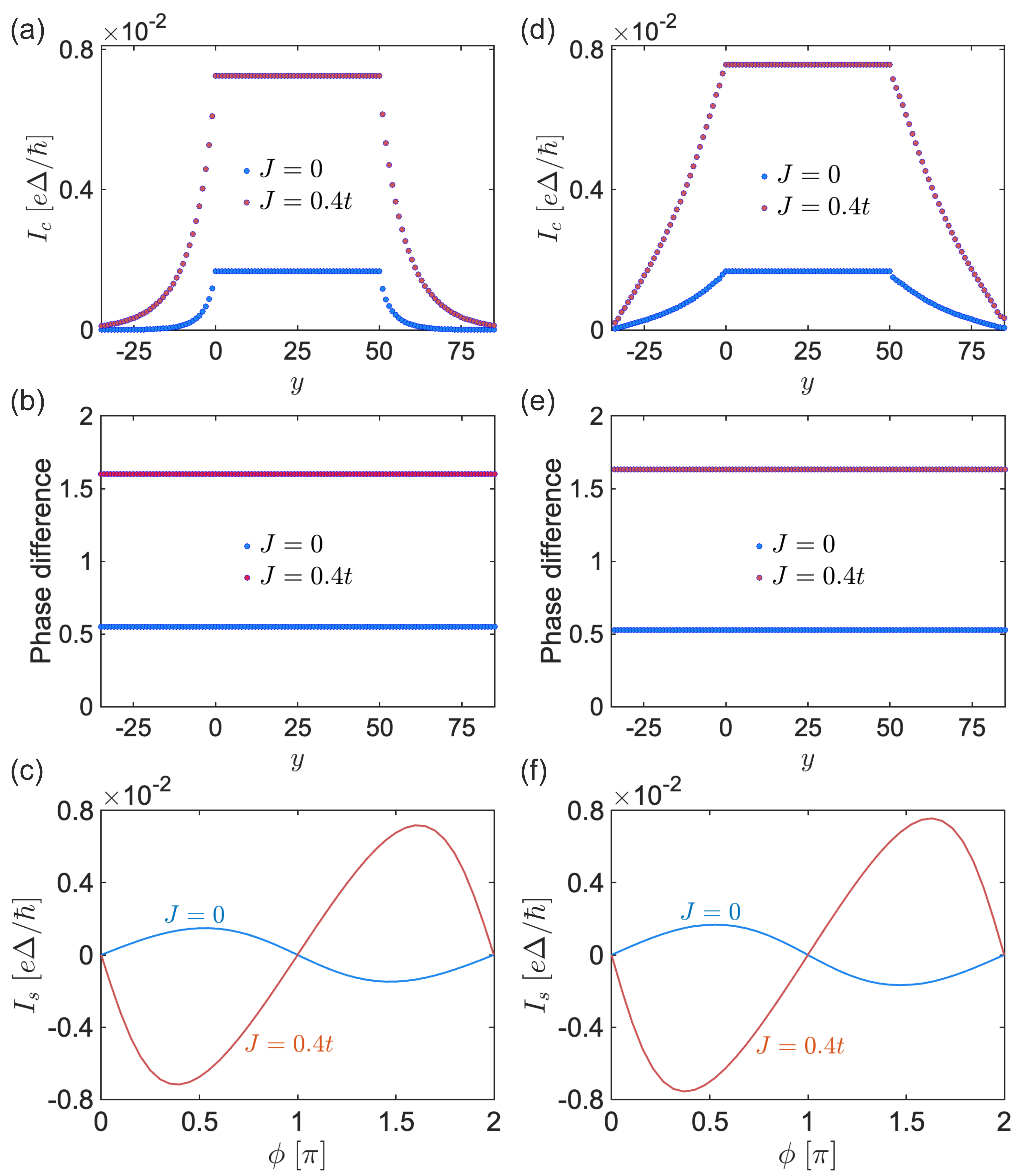}
    \caption{(a) Maximum supercurrent $I_c$ and (b) corresponding phase position $\phi_I^*$ (in units of $\pi$) as functions of $y$ in the junction for $J=0$ (blue dots) and $0.4t$ (red dots). (c) CPR $J_s(\phi)$ calculated at the junction center for $J=0$ (blue line) and $0.4t$ (red line). In (a)-(c), only the magnitude of the superconducting order parameters is determined self-consistently, while the phases are fixed at $0$ and $\phi$ in the two superconducting leads.  
    (d)-(f) Same as (a)-(c), but with both the magnitude and phases of the order parameters determined self-consistently, except the outermost five layers at both ends of the junction, where the phases remain fixed at $0$ and $\phi$, respectively. Other parameters: $\mu_S=2t$, $\mu_{\text{AFM}}=0.1t$, $N_L=50$, $U=0.65t$ (yielding $\Delta=0.023t$ and $T_ck_B=0.013t$), and $T=0.1T_c$.}
    \label{fig:Js_y_and_Js_phi}
\end{figure}
We calculate the supercurrent using the self-consistent obtained superconducting order parameter in the formula in Eq.~\eqref{eq:kernel_fun}. 
For illustration, Fig.~\ref{fig:Js_y_and_Js_phi} presents the results for $J=0$ (blue) and $0.4t$ (red), with temperature $T=0.1T_c$, $N_L=50$, $\mu_{\text{AFM}}=0.1t$ and all other parameters the same as Fig.~\ref{fig:Delta_y_phi}. 
We find that the supercurrent $I_s$ is constant in the cAFM region ($1\leq y \leq N_L$), due to particle conservation (see Figs.~\ref{fig:Js_y_and_Js_phi}(a) and (b) for the maximum superrent and its phase position in the $\phi$-axis). In contrast, the supercurrent decreases monotonically when moving into the superconducting leads. Moreover, the supercurrent is significantly enhanced by the cAFM ($J=0.4t$) compared to the nonmagnetic case ($J=0$). Specifically, in the cAFM region, the maximum supercurrent increases from {$I_c\approx 0.16\times 10^{-2} e\Delta/\hbar$ for $J=0$ to $I_c\approx 0.75\times 10^{-2} e\Delta/\hbar$ for $J=0.4t$}. The current-phase relations (CPRs) for these two cases are displayed in Fig.~\ref{fig:Js_y_and_Js_phi}(c), demonstrating a $0$-junction in the nonmagnetic case, whereas a $\pi$-junction when $|\mu_{\text{AFM}}|<|J|$. 
These results remain consistent, regardless of whether the phase of the order parameter across the junction is also determined self-consistently (comparing the left and right columns in Fig.~\ref{fig:Js_y_and_Js_phi}).  
Thus, to reduce computational costs, it is sufficient to compute only the magnitude of the order parameter self-consistently while keeping its phase fixed in the superconducting leads.




%

\end{document}